\documentclass[iop]{emulateapj}
\usepackage{natbib}

\shorttitle{Nature of HD\,85567} \shortauthors{S.A. Khokhlov et al.}

\begin{document}

\title{Towards Understanding The B[e] Phenomenon: VI. Nature and Spectral Variations of HD\,85567.}

\author{S.~A.~Khokhlov$^{1,2}$}
\affil{$^1$Physico-Technical Department, Al Farabi Kazakh National University, Al Farabi Av., 71, 050040, Almaty, Kazakhstan}

\author{A.~S.~Miroshnichenko$^{2,3}$}
\affil{$^2$Department of Physics and Astronomy, University of North
Carolina at Greensboro, P.O. Box 26170, Greensboro, NC 27402--6170, USA}
\affil{$^3$National center of space exploration and technologies, Almaty, Kazakhstan}

\author{R.~Mennickent$^4$, M.~Cabezas$^4$}
\affil{$^4$Department of Astronomy, Universidad de Concepci\'on
Barrio Universitario, Av. Esteban Iturra, Casilla 160--C, Concepci\'on, Chile}

\author{Z.~Zh.~Zhanabaev$^1$}
\affil{$^1$Physico-Technical Department, Al Farabi Kazakh National University, Al Farabi Av., 71, 050038, Almaty, Kazakhstan}

\author{D.~E.~Reichart$^5$, K.~M.~Ivarsen$^5$, J.~B.~Haislip$^5$, M.~C.~Nysewander$^5$, A.~P., LaCluyze$^5$}
\affil{$^5$Department of Physics and Astronomy, University of
North Carolina at Chapel Hill, Campus Box 3255, Chapel Hill, NC 27599, USA}

\altaffiltext{1}{This paper is partly based on observations obtained
at the 1.5\,m telescope of the Cerro Tololo Interamerican Observatory under CNTAC proposal 2012A--016.}

\begin{abstract}
We report the results of high-resolution ($R \sim 80,000$) spectroscopic observations of the emission-line object HD\,85567, which has been classified as a
FS\,CMa type object or a pre-main-sequence star. The main goal is to put more constraints on the object's fundamental parameters
as well as on its nature and evolutionary state. Absorption lines in the spectrum of HD\,85567 were found similar to those of mid B--type dwarfs and correspond
to the following fundamental parameters: T$_{\rm eff} = 15000\pm500$ K, $v \sin i = 31\pm3$ km\,s$^{-1}$, $\log g \sim 4.0$. The interstellar  extinction,
A$_V$ $= 0.50\pm0.02$ mag, was measured using the strengths of some diffuse interstellar bands.
We also obtained $UBV(RI)_{\rm c}$ images of a $10\arcmin \times 10\arcmin$ region around the object. Photometry of projectionally close stars was
used to derive an interstellar extinction law  in this direction and resulted in a distance of 1300$\pm$100 pc to the object
and a luminosity of $\log$ L/L$_\odot = 3.3\pm0.2$. We found no significant radial velocity variations of the absorption lines in the spectra of HD\,85567 obtained
during two month-long periods of time in 2012 and 2015. Our analysis of the spectroscopic and photometric data available for the star led us to a conclusion
that it cannot be a pre-main-sequence Herbig Ae/Be star. We argue that the circumstellar gas and dust were produced during the object's evolution as most
likely a binary system, which contains an undetected secondary component and is unlikely to be a merger product.
\end{abstract}


\section{Introduction} \label{intro}

The B[e] phenomenon is defined as the presence of forbidden line emission (in addition to permitted line emission) and infrared (IR) excess due to circumstellar dust in the spectra of B--type stars. It was discovered by \citet{as76} and is observed in stars with a wide range of masses and evolutionary states. Despite a strong progress in understanding of these complex objects, many of them are still poorly studied. In particular, half of the objects from the initial list of 65 Galactic stars that show the B[e] phenomenon were not identified as members of any group with known evolutionary state. These objects were called unclassified by \citet{L98}.

 \citet{m07} suggested that most unclassified objects were neither pre-main-sequence Herbig Ae/Be (HAeBe) stars nor supergiants but rather binary systems and separated them into a new group of FS\,CMa type objects. Limitation of the mass range and evolutionary states for the group was based on the strength of the emission-line spectra and shape of the IR excess radiation (lack of far-IR emission). \citet{m07a} expanded the FS\,CMa group with 10 newly discovered objects found in the {\it IRAS} database by cross-identification with catalogs of optical positions. \citet{m11} reported $\sim$20 more candidates found in the NOMAD catalog \citep{z05} using optical and near-IR photometric criteria. Although the range of possible interpretations for the group objects properties was limited, binarity has been detected only in $\sim$30\% of the group members. At the same time, information about some group objects is not detailed enough that allows different researchers reach contradictory conclusions on their  nature and evolutionary state.

This paper is devoted to a study of HD\,85567, a bright ($V \sim$ 8.6 mag) star that is located in the southern hemisphere in the Carina constellation.
Its emission-line spectrum was first reported by \citet{1976ApJS...30..491H}. The star was included in a catalog of Be stars by \citet{1982IAUS...98..261J}. \citet{1990AcApS..10..154H} and \citet{1992A&AS...96..625O} reported a strong IR excess detected from the star by {\it IRAS}. \citet{1994A&AS..104..315T} included HD\,85567 in their catalog of HAeBe candidate stars. \citet{L98} in their classification of B[e]--type stars also suggested that HD\,85567 was a HAeBe star. This conclusion was adopted in a number of other studies \citep[e.g.,][]{1998A&A...330..145V}.

Nevertheless, \citet{2001A&A...371..600M} showed that this star exhibited a much weaker far-IR excess compared to genuine HAeBe stars. Additionally these authors speculated that it was a binary system because of a complex structure of the Balmer line profiles and a large difference between radial velocities (RV) of absorption and emission in its spectrum. A few years later \citet{2006MNRAS.367..737B} detected signs of the object's binarity by spectro-astrometry.   \citet{m07} included it into the
FS\,CMa group. However, recent studies kept considering HD\,85567 a pre-main-sequence star \citep[e.g.,][]{2003AJ....126.2971V,2012A&A...538A.101V,2014MNRAS.445.3723I,2015MNRAS.453..976F}.
Moreover the interferometric observations  of \citet{2013A&A...558A.116W} did not  detect any binary signature,  although they did not exclude the possibility that a faint companion could still escape detection. These authors concluded that HD\,85567 is a young stellar object with an optically-thick gaseous disk within a larger dusty disk that is being photo-evaporated from the outer edge. Therefore there are two conflicting opinions on the object's nature and evolutionary state.

\begin{table}[!h]
\caption[]{Fundamental parameters of HD\,85567 from the literature}\label{t1}
\begin{center}
\begin{tabular}{lllllc}
\hline\noalign{\smallskip}
SpT   & T$_{\rm eff}$  &  D, kpc  & $\log$ L/L$_\odot$  & A$_V$         & Ref.\\
\noalign{\smallskip}\hline\noalign{\smallskip}
B2    & 19000   &   1.5$\pm$0.5  &   4.0$\pm$0.3         & 1.25              & 1   \\
B2    & 22000   &   1.5                 &   4.66                      & 2.23              & 2   \\
B5    & 15200   &   $\ge$0.48       &   $\ge$2.54             & 0.71              & 3   \\
B2    & 22000   &   1.5$\pm$0.5  &   4.17$\pm$0.16     & 1.1$\pm$0.1 & 4   \\
         &  13000$\pm$500 & 0.9$^{+0.2}_{-0.1}$  &  3.13$^{+0.46}_{-0.45}$ & 0.89$^{+0.03}_{-0.02}$ & 5 \\
 B4--5  &  15000$\pm$500  & 1.3$\pm$0.1 &  3.3$\pm$0.2  &  0.50$\pm$0.02 &  6\\
\noalign{\smallskip}\hline
\end{tabular}
\end{center}
\begin{list}{}
\item Column information: (1) -- spectral type, (2) -- effective temperature in degrees Kelvin,
(3) -- distance in kiloparsecs, (4) -- logarithm of the luminosity in solar units,
(5) -- visual interstellar extinction, and (6) -- reference.
\item References: 1 -- \citet{2001A&A...371..600M}, 2 -- \citet{2006ApJ...653..657M},
3 -- \citet{2008A&A...484..225M}, 4 -- \citet{2012A&A...538A.101V},  5 -- \citet{2015MNRAS.453..976F},
 6 --  this paper.
\end{list}
\end{table}

Fundamental parameters of HD\,85567 found in the literature are summarized in Table \ref{t1}. We note here that the distance quoted by \citet{1998A&A...330..145V}
came from the original version of the {\it HIPPARCOS} catalog of parallaxes \citep[1.00$\pm$0.69,][]{1997ESASP1200.....E}. All the other distances shown in
Table\,\ref{t1} refer to \citet{2001A&A...371..600M}. The updated {\it HIPPARCOS} parallax \citep[0.47$\pm$0.51,][]{2007A&A...474..653V} gives an unreliable distance.

In this paper we use multicolor photometry of stars in the vicinity of HD\,85567 and its high-resolution spectra to refine its fundamental parameters and re-investigate its nature and evolutionary state. The observations are described in Sect. \ref{observations}, analysis of the spectrum and the radial velocity curve is presented in Sect. \ref{analysis}, suggestions about the system's nature are discussed in Sect. \ref{discussion}, and conclusions are summarized in Sect. \ref{conclusions}.

\section{Observations}\label{observations}

\begin{figure*}[!htb]
\begin{center}
\begin{tabular}{c}
\includegraphics[width=18cm, height=6cm]{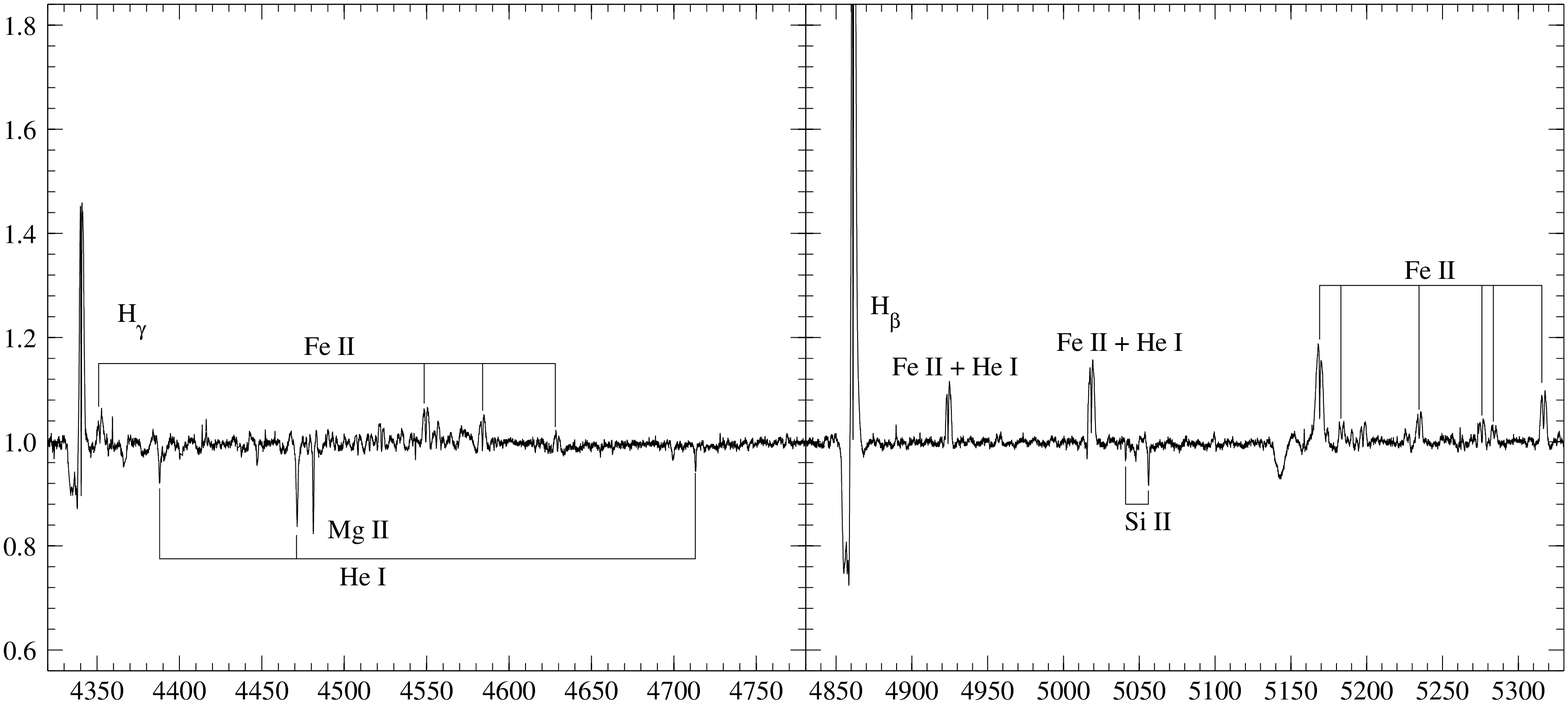}\\
\includegraphics[width=18cm, height=6cm]{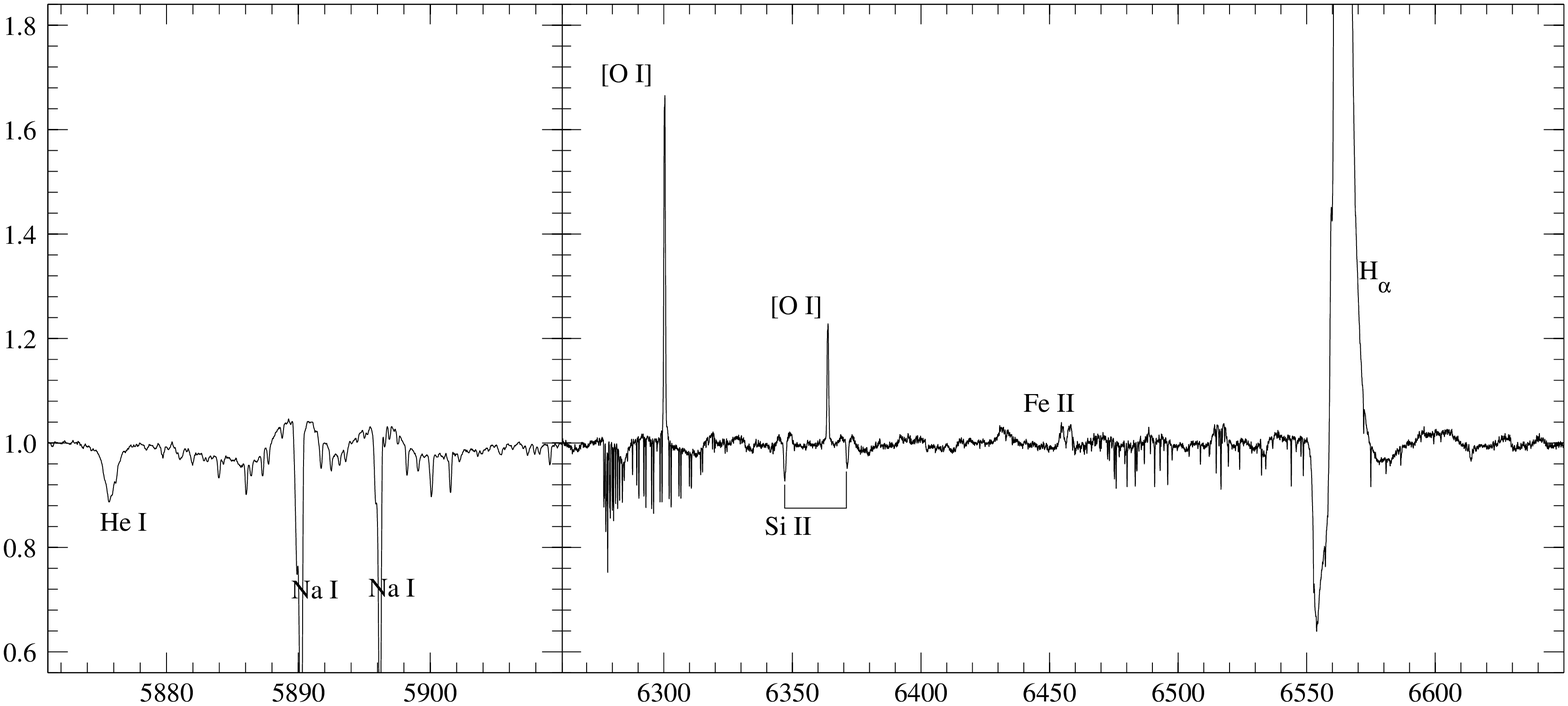}\\
\includegraphics[width=18cm, height=6cm]{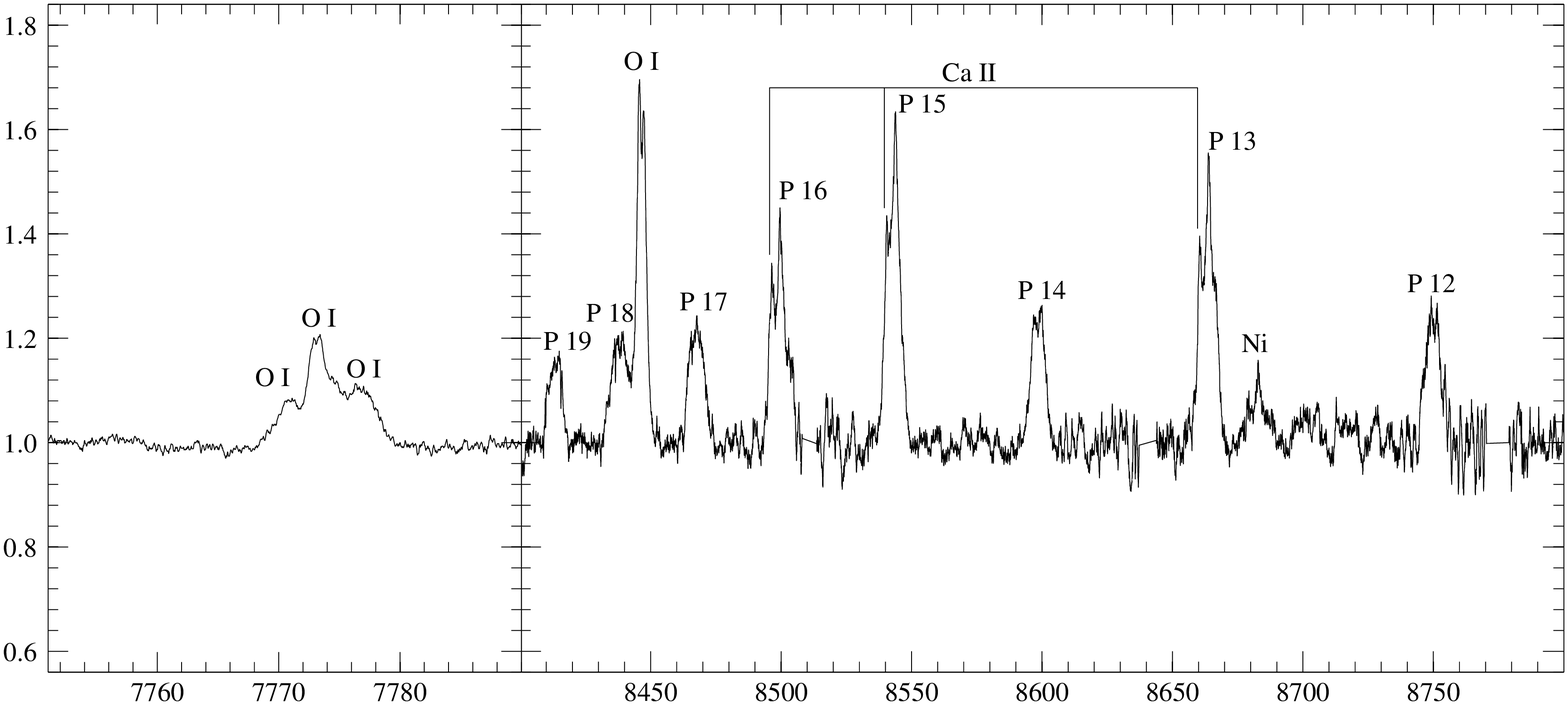}\\
\end{tabular}
\end{center}
\caption {The most informative parts of a high-resolution HIRON spectrum of HD\,85567 ($R = 80,000$).
The intensity is normalized to the continuum, the wavelength scale is heliocentric.}
\label{f1}
\end{figure*}

Optical high-resolution spectroscopic observations of HD\,85567 were obtained between March 3 and April 5, 2012 (5 spectra) and between April 18 and May 23, 2015
(10 spectra) at a 1.5\,m telescope of CTIO with the HIRON spectrograph \citep{2013PASP..125.1336T} using an image slicer. {\ The MJD of all spectroscopic observations presented in Table.\,\ref{f4}.}
The spectral resolving power was $R = 80,000$. The 2012 spectra were obtained in a spectral range of 4000--9000 \AA, and the 2015 spectra were obtained in a spectral range of 4500--9000 \AA.  All the spectra were obtained with an exposure time of $\sim$1200 s. Both data sets are characterized by an average signal-to-noise ratio of $\sim$100. Data reduction was done with a pipeline\footnote{http://www.ctio.noao.edu/noao/sites/default/files/telescopes/\\smarts/tele15/chireduce.pdf
}  using standard IRAF\footnote{IRAF is distributed by the National Optical Astronomy Observatory, which is operated by the Association of Universities for Research in Astronomy (AURA) under a cooperative agreement with the National Science Foundation.} tasks and included  bias subtraction, flat--fielding, extracting one-dimensional spectra, normalizing each order to the continuum, wavelength calibration using a Th-Ar lamp exposures, and correcting wavelengths for the Earth translational motion. An example of our spectrum of HD\,85567 with a representative set of spectral lines is presented in Fig.\,\ref{f1}.

\begin{table*}[!htb]
\caption[] {Photometry of stars in the vicinity of HD\,85567}\label{t2}
\begin{center}
\begin{tabular}{cccccccclrr}
\hline\noalign{\smallskip}
1        &2         &3      &4      & 5    & 6   &  7   &  8   &  9 & 10&11\\
\noalign{\smallskip}\hline\noalign{\smallskip}
1   & 9 49 58 &$-$61 00 14 & 10.91  & 0.15 & 0.64 & 0.35 & 0.76 & F5 {\sc iii}  & 0.45 & 0.5 \\
2   & 9 50 09 &$-$61 03 01 & 10.60  & 0.82 & 1.15 & 0.57 &1.18  & G5 {\sc iii}  & 0.50 & 0.7 \\
3   & 9 51 08 &$-$60 54 35 & 11.68  &  --    & 0.41 & 0.23 & 0.57 & F2 {\sc iv}   & 0.25 & 0.5 \\
4   & 9 50 33 &$-$60 57 32 & 13.82  & 0.44 & 0.89 & 0.46 & 1.09 & G0 {\sc iii}  & 0.85 & 2.0 \\
5   & 9 50 24 &$-$60 59 08 & 12.53  & 1.10 & 1.29 & 0.62 & 1.31 & G8 {\sc iii} & 0.80 & 1.5 \\
6   & 9 50 12 &$-$60 59 23 & 12.29  & 2.55 & 1.65 & 0.87 & 1.81 & --      & --    & --   \\
7   & 9 50 04 &$-$60 59 16 & 13.61  & 0.11 & 0.65 & 0.33 & 0.78 & F5 {\sc iv}  & 0.45 & 0.8 \\
8   & 9 50 06 &$-$60 58 28 & 14.07  & 0.30 & 0.74 & 0.42 & 0.88 & F8 {\sc iv}  & 0.45 & 0.8 \\
9  & 9 50 09 &$-$60 58 05 & 13.17  & 1.00 & 1.20 & 0.63 & 1.31 & G5 {\sc iii}  & 1.05 & 1.7 \\
10  & 9 50 06 &$-$60 57 05 & 13.96  & 0.10 & 0.66 & 0.36 & 0.82 & F2 {\sc iii}  & 0.85 & 1.8 \\
11  & 9 50 14 &$-$60 56 59 & 13.38  &   --   & 1.30 & 0.64 & 1.34 & G5 {\sc iii}  & 0.85 & 2.0 \\
12  & 9 50 21 &$-$60 56 48 & 13.68  & 1.19 & 1.31 & 0.64 & 1.34 & G8 {\sc iii}  & 0.95 & 2.3 \\
13  & 9 50 45 &$-$60 58 07 & 12.96  & 1.22 & 1.87 & 1.72 & 4.16 & --      & --    & --   \\
14  & 9 50 27 &$-$60 55 21 & 13.26  & 0.10 & 0.39 & 0.21 & 0.51 & F2 {\sc v}    & 0.20 & 0.9 \\
\noalign{\smallskip}\hline
\end{tabular}
\end{center}
\begin{list}{}
\item Column information: (1) -- sequential number or a star in the vicinity of HD\,85567, (2) -- RAJ2000, (3) -- DECJ2000
 (4) -- V, (5) -- $U-B$, (6) -- $B-V$, (7) -- $V-R$, (8) -- $V-I$, (9) -- assigned MK type, (10) -- A$_V$, (11) -- distance D in kpc.
The coordinates are taken from the NOMAD catalog \citep{z05}. For determination of MK types see Sect. \ref{extinction}.
Stars Nos. 6, 13 are very red and probably have late spectral types, so that luminosity type determination is uncertain.
\end{list}
\end{table*}

\begin{figure}[!h]
\begin{center}
\includegraphics[width=8cm, height=8cm]{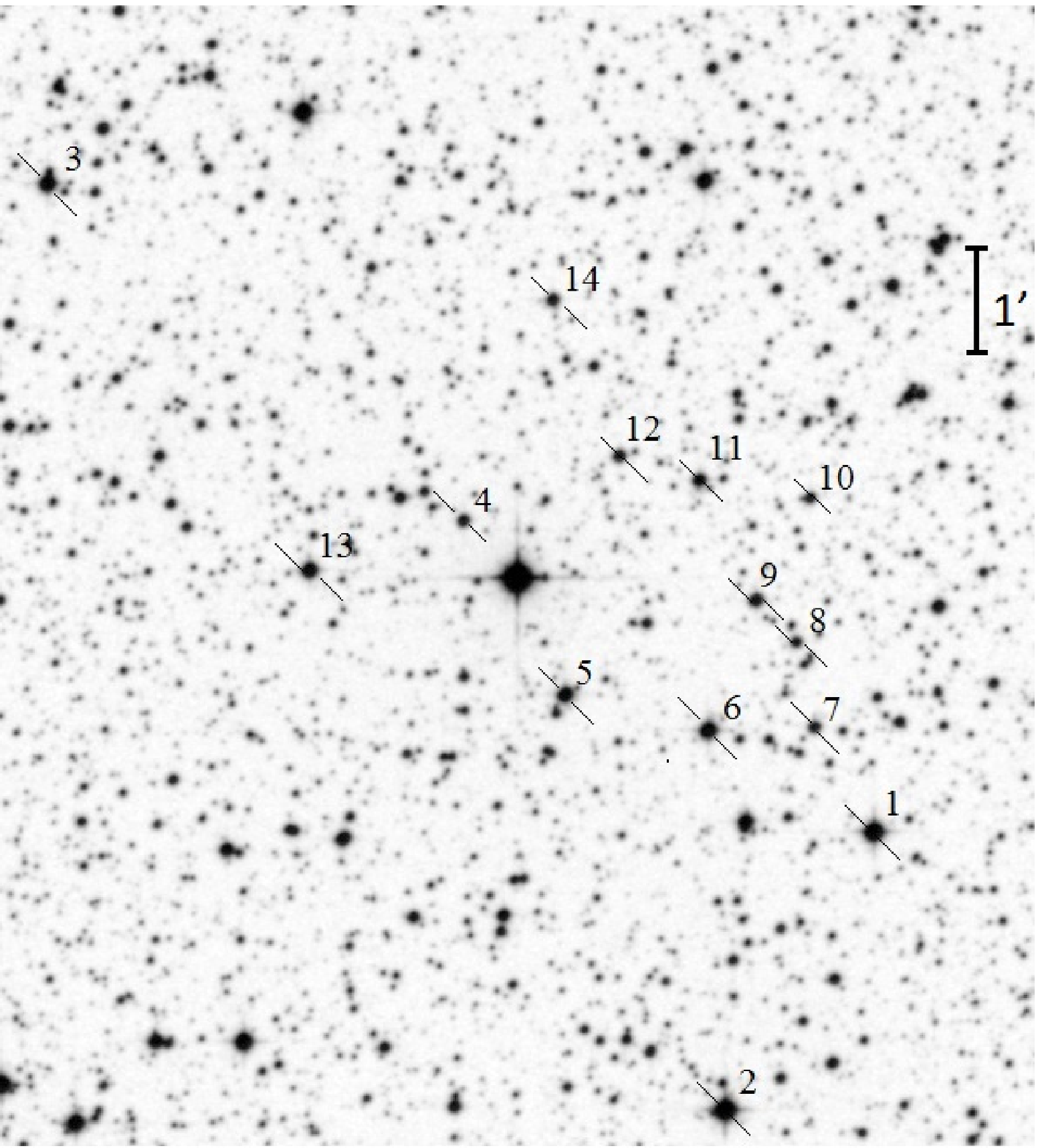}
\end{center}
\caption{{The $10\arcmin \times 10\arcmin$ field around HD\,85567 from the Digital Sky Survey. The object is the brightest star in the middle of the field. Stars,
whose brightness was measured and is presented in Table\,\ref{t2}, are marked with numbers. The North direction is up, and East is to the left.}
\label{f2}}
\end{figure}

$UBV(RI_{\rm c})$ data in the Johnson-Cousins photometric system of a $10\arcmin \times 10\arcmin$ field around HD\,85567
and some projectionally close fields were obtained in February--April and November 2015 at three PROMPT robotic telescopes \citep{2005NCimC..28..767R}
located at CTIO. The images were bias subtracted, flat fielded, and dark corrected. The brightness of the stars shown in Fig.\,\ref{f2} as well as of several stars from the other images was measured using IRAF task {\it imexamine}  with apertures based on seeing (typical radius $\sim$10 pixels or $\sim 6\arcsec$).
Transformation coefficients between the instrumental and standard photometric system were determined from observations of several fields containing standard stars from \citet{1992AJ....104..340L}. The measurement results in the standard system are presented in Table \ref{t2}.

\section{Data Analysis}\label{analysis}
 We determined both the absorption and emission lines RV by fitting the line profiles to a Gaussian. Particularly in the case of double and triple peaked emission lines each peak was fitted to a Gaussian. The emission peaks in the [O {\sc i}] lines are so close together that we determined the RV by fitting the symmetric part of the line profile wings to a single Gaussian (see Table\,\ref{t4}). The equivalent widths (EW) were measured by direct integration above the local continuum.

\subsection{Absorption lines} \label{absorptions}

The spectrum of HD\,85567 shows a number of absorption lines typical for a B--type star atmosphere as well as those originated in the interstellar and circumstellar medium.
The former group includes the He {\sc i} lines (4471, 4713, 5875, 6678 \AA), Mg {\sc ii} 4481 \AA, Si {\sc ii} (5041, 5056, 6347, 6371 \AA). RVs of some of these lines are shown in Table\,\ref{t3}. Their average value is 0.6$\pm$3 km\,s$^{-1}$ except for the spectrum taken on 03/30/2012, which seems to be affected by slight wavelength calibration problems. The scatter of the average RV is most likely due to a small number of detected absorption lines and contamination by the circumstellar material (see also Sect.\,\ref{binary} for a discussion of the orbital motion as a reason for the RV scatter). There is virtually no difference with the earlier result by \citet{2001A&A...371..600M}, 0$\pm$2 km\,s$^{-1}$ based on a single spectrum obtained in 2000.

The Mg {\sc ii} 4481 \AA\ and Si {\sc ii} 6347 and 6371 \AA\ lines have weak emission components ($\sim$0.01--0.02 of the local continuum intensity due to circumstellar contribution, see Fig.\,\ref{f1}) that make the lines slightly weaker compared to those of normal (with no line emission) stars. Therefore, EW of these lines may not be directly used as the effective temperature (T$_{\rm eff}$) indicators for this object. However, EW ratios of various absorption lines are less affected by emission and work reasonably well for T$_{\rm eff}$ determination. \citet{2015ApJ...809..129M} showed that the EW ratio of the He {\sc i} 4471 \AA\ and the Mg {\sc ii} 4481 \AA\ is a good T$_{\rm eff}$ indicator for some Be stars and the FS\,CMa object MWC\,728 (see their Fig.\,3). This ratio in the spectrum of HD\,85567 is equal to 1.3$\pm$0.2 that corresponds to T$_{\rm eff}=$13500$\pm$1000 K and a spectral type B5 \citep{gray94}.

\begin{figure}[!h]
\begin{center}
\includegraphics[width=9.0cm, height=6.0cm]{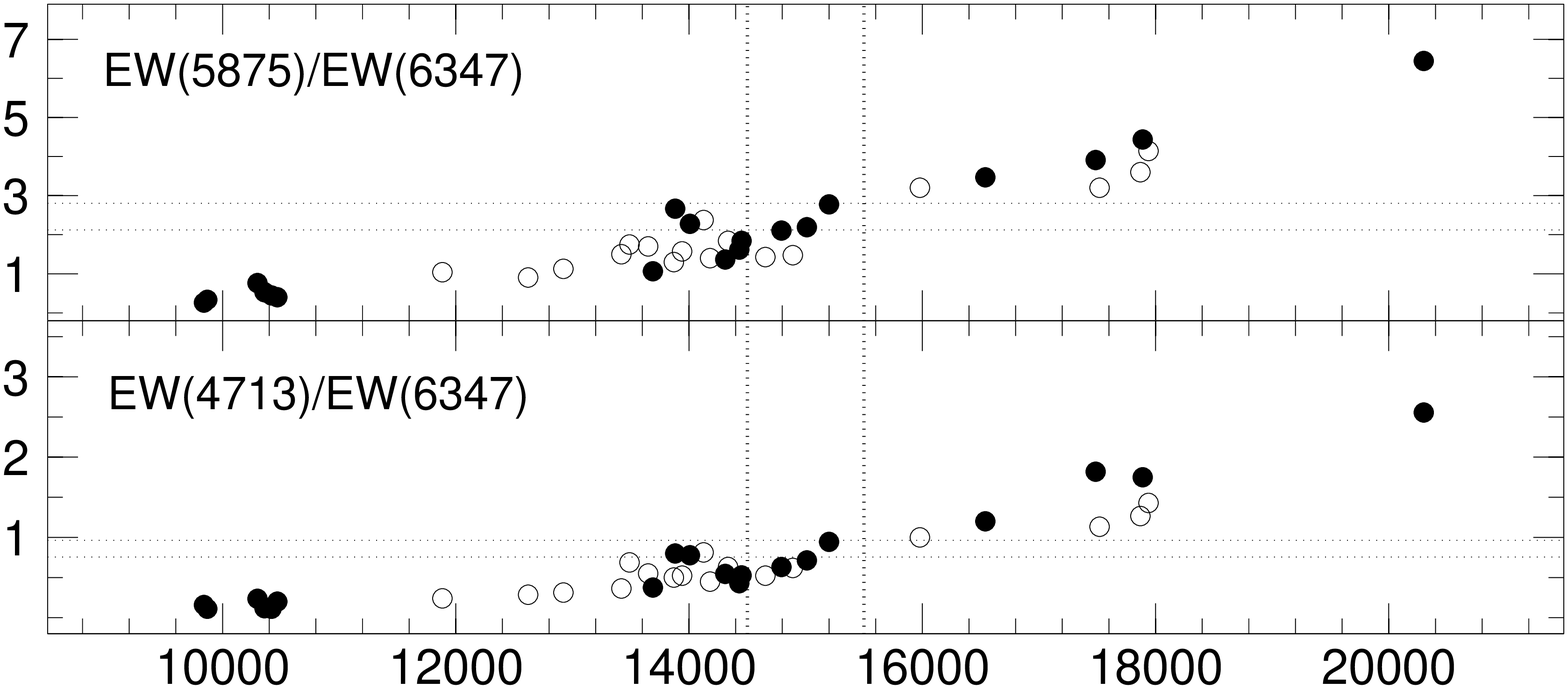}
\end{center}
\caption{{$T_{\rm eff}$ relationship with the EW ratio the He {\sc i} 4713 \AA\ and the Si {\sc ii} 6347\AA\ lines and EW ratio  He {\sc i} 5875 \AA\ and the Si {\sc ii} 6347\AA\ lines. Circles represent data for normal B--type stars (filled— -- OHP, open— --TCO). Temperature determination were collected from various papers \citep{2009A&A...501..297Z, 2012A&A...537A.120Z}. The horizontal dashed lines show ranges of EW  variations detected in the spectrum of HD\,85567, and the vertical dashed lines show most likely boundaries of the $T_{\rm eff}$ for HD\,85567.}
\label{f3}}
\end{figure}

\begin{figure}[!h]
\begin{center}
\includegraphics[width=9.0cm, height=6cm]{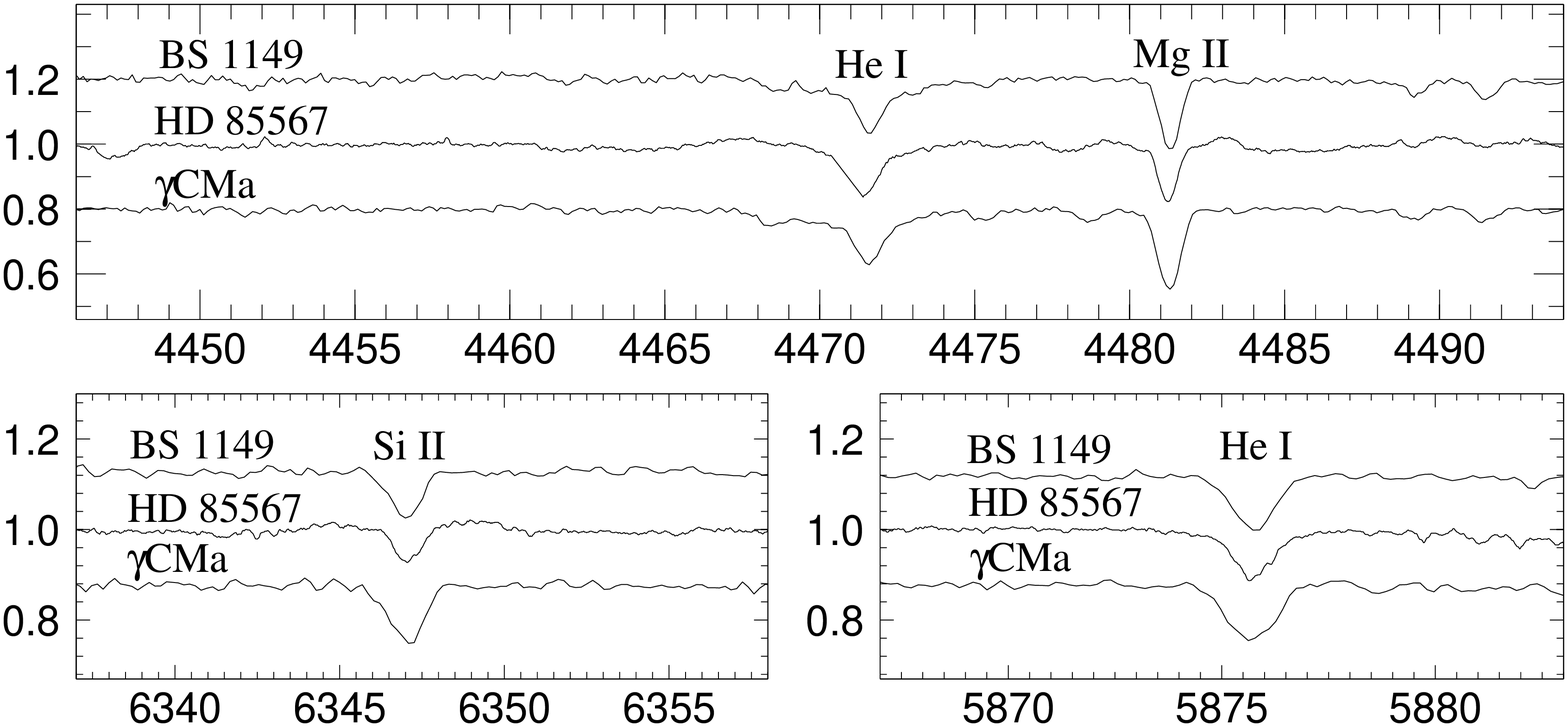}
\end{center}
\caption{{The He {\sc i} (4471 and 5875 \AA), the Mg {\sc ii} 4481 \AA, and the Si {\sc ii} 6347 \AA\  lines in the spectrum of HD\,85567 taken on 2012 March 19 are shown in the middle of each panel. An {\it ELODIE} spectrum of BS\,1149 (T$_{\rm eff}$ = 14310 K, $v \sin\,i = 33$ km\,s$^{-1}$) and a TCO spectrum of $\gamma$ CMa (T$_{\rm eff}$ = 13690 K,  $v \sin\,i = 25$ km\,s$^{-1}$) are shown for comparison. The heliocentric wavelengths are shown in \AA, the intensity is normalized to the nearby continuum. The spectra are shifted by 0.1\,I$_{\rm cont}$ with respect to each other for an easier comparison.}
\label{f4}}
\end{figure}

We also employed the EW ratio of the He {\sc i} 4713 \AA\ and the Si {\sc ii} 6347 \AA\ lines and that of the He {\sc i} 5875 \AA\ and the Si {\sc ii} 6347 \AA\ as additional
T$_{\rm eff}$ indicators to verify the above result. We used spectra of normal B--type stars which were obtained at the 0.81\,m telescope of the Three College Observatory of the University of North Carolina at Greensboro with an \'echelle spectrograph that provides $R \sim$10,000 and those obtained at the 1.93\,m telescope of the Observatoire de Haute Provence with the spectrograph {\it ELODIE} \citep[$R \sim$ 42000,][]{2004PASP..116..693M}. Using the relationships of the lines EW and $T_{\rm eff}$ shown in Fig.\,\ref{f3}, we estimated the object's $T_{\rm eff}=$15000$\pm$500~K. Comparing spectra of MWC\,728 \citep[no Fe {\sc ii} lines in emission, T$_{\rm eff}=$14000$\pm$1000 K,][]{2015ApJ...809..129M} and HD\,85567 and taking into account that the latter exhibits Fe {\sc ii} emission lines, we favor the higher T$_{\rm eff}$ for it. We determined a projected rotational velocity of $v \sin\,i = 31\pm3$ km\,s$^{-1}$ from profiles of the He {\sc i} (4713, 5875, 6678 \AA) and Si {\sc ii} (5056, 6347, 6371 \AA) lines using Fourier transform. In Fig.\,\ref{f4} we compare the spectrum of HD\,85567 with those of normal B--type stars having similar rotation velocities.

The latter group of absorption lines contains diffuse interstellar bands (DIBs), such as 5780, 5797, and 6613 \AA, and the Na {\sc i} D--lines at 5889 and 5895 \AA. Each of the Na {\sc i} D--lines exhibits two peaks (V$_{\rm blue}= -2.0\pm$0.6 km\,s$^{-1}$, V$_{\rm red} = 13.7\pm$0.1 km\,s$^{-1}$). The blue-shifted peak most likely forms in the circumstellar material. \citet{2001A&A...371..600M} determined the interstellar extinction for HD\,85567 as $A_{V} \sim 0.5$ mag using the 5780 and 5797 \AA\ DIB strengths. The DIB strengths in our new spectra are essentially the same. The average EW of the 5780 \AA\ DIB is 0.08 \AA\ that corresponds to A$_{V}$ = 0.50$\pm$0.02 mag according to a calibration by \citet{1993ApJ...407..142H}.

\begin{table}[!h]
\caption[] {RVs of some absorption lines in the spectra of HD\,85567}\label{t3}
\begin{center}
\begin{tabular}{crrrrrr}
\hline\noalign{\smallskip}
date   & Mg {\sc ii} & He {\sc i}  &He {\sc i} &Si {\sc ii}& Si {\sc ii}    & He {\sc i}  \\
          & 4481         & 4713        & 5875       & 6347    & 6371          & 6678         \\
\noalign{\smallskip}\hline\noalign{\smallskip}
03/03/2012  &$-$0.4  & 6.3     &3.1     & 2.0   &$-$0.8 & 0.2   \\
03/19/2012  & 0.4      & 1.3      &4.6     &$-$1.9 &$-$2.8 &$-$2.7 \\
03/23/2012  & 0.0     & 0.9       &4.1     & 0.5   & 0.8   &$-$2.4 \\
03/30/2012  & $-$     &$-$4.6  &$-$10.7 &$-$2.8 & 0.5   &$-$0.6 \\
04/05/2012  & 3.1     & 4.9      &$-$0.9  & 1.6   & 2.2   &$-$3.1 \\
04/18/2015  & &                      & 3.6    & 0.2   & 0.5   & 2.2   \\
04/20/2015  & &                     & 3.9    &$-$0.5 & 2.8   & 0.1   \\
04/23/2015  & &                     & 3.9    &$-$0.6 & 1.9   &$-$3.0   \\
04/25/2015  & &                     & 4.1    &$-$2.7 & 0.0   &$-$0.3 \\
04/27/2015  & &                     & 3.6    &$-$1.4 & 0.3   & 0.0   \\
04/29/2015  & &                     & 3.6    &$-$2.2 & 1.9   & 0.7   \\
05/01/2015  & &                     & 4.1    &$-$1.4 & 1.3   &  $-$     \\
05/02/2015  & &                     & 3.9    & 0.3   & 1.7   & 0.4   \\
05/05/2015  & &                     & 4.6    &$-$2.4 & 2.4   & 1.3   \\
05/23/2015  & &                     & 3.9    & 0.0   & 1.1   &$-$0.6 \\
\noalign{\smallskip}\hline\noalign{\smallskip}
Average  & $-$0.8 & 4.4         &  3.6   & $-$0.6 &  1.0  &  $-$0.6 \\
r.m.s.       &      1.2 & 2.3         & 1.1    & 1.3   & 1.1   & 1.3   \\
\noalign{\smallskip}\hline
\end{tabular}
\end{center}
\begin{list}{}
\item Heliocentric RVs (in km\,s$^{-1}$) of some absorption lines in the spectrum of HD\,85567. The last raw shows r.m.s. standard deviations for the
average RV values. The RVs measured in the spectrum taken on 03/30/2012 are excluded from averaging (see Sect.\,\ref{discussion}).
The He {\sc i} 4471 \AA\ and the Mg {\sc ii} 4482 \AA\ lines were out of the spectral range detected in 2015.
The He {\sc i} 6678 \AA\ line in the spectrum taken on 05/01/2015 is too noisy for a reliable RV measurement.
\item Julian dates of the spectroscopic observations are shown in Table\,\ref{t4}.
\item Rest wavelengths of the lines in this and other Tables were taken from \citet{1993BICDS..43....7C}
\end{list}
\end{table}

\subsection{Emission lines}\label{emission}

\begin{figure} [!htb]
\begin{center}
\includegraphics[width=8.5cm, height=6cm]{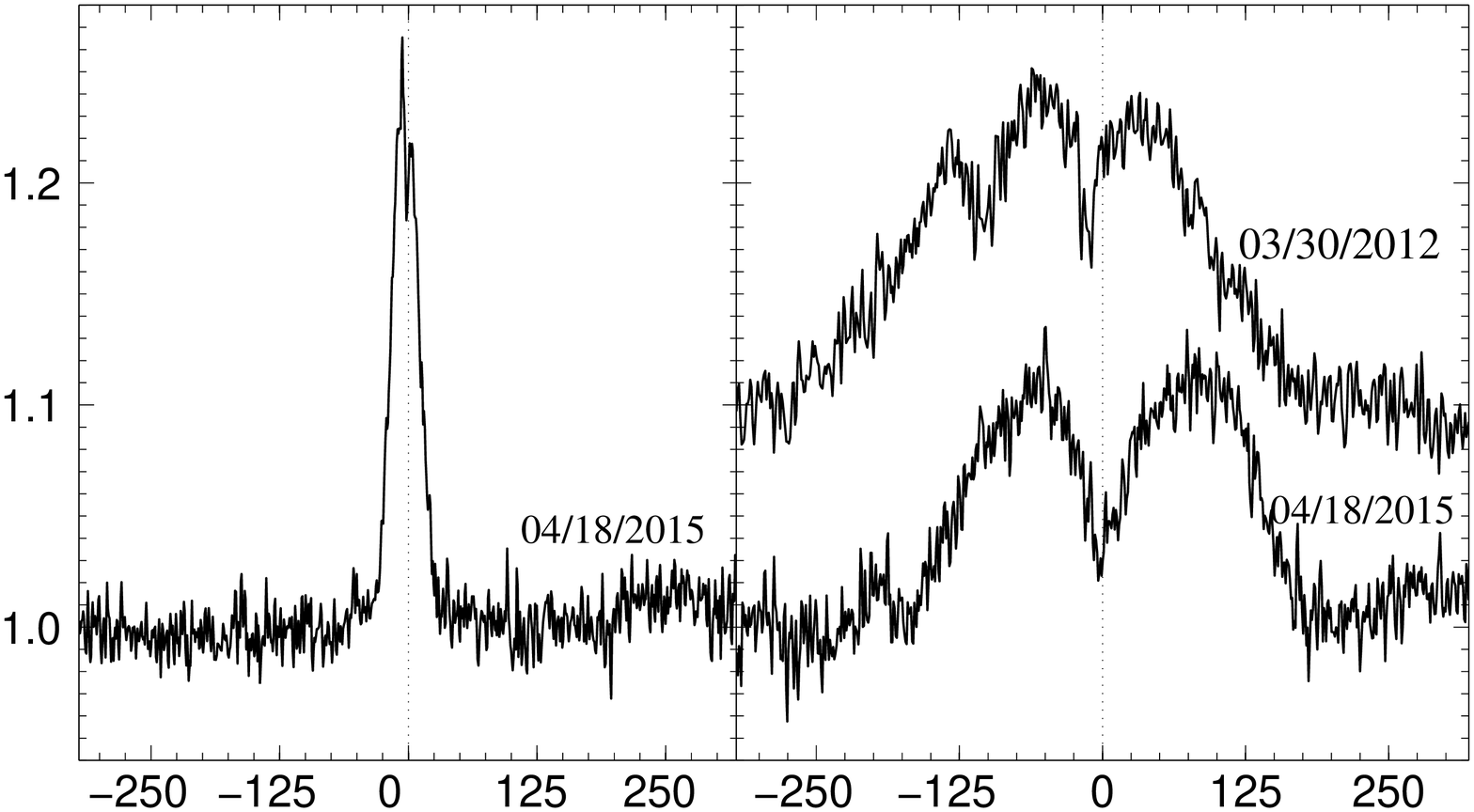}
\end{center}
\caption{
 The [O {\sc i}] 6364 \AA\ (left panel) and the Fe {\sc ii} 5169 \AA\ (right panel) emission lines in the spectrum of HD\,85567.
The right panel shows an example of the Fe {\sc ii} line profile variations. Both the intensity and radial velocity are in the same units as in Fig.\,\ref{f4}.
\label{f5}
}
\end{figure}

\begin{figure} [!htb]
\begin{center}
\includegraphics[width=8.5cm, height=6cm]{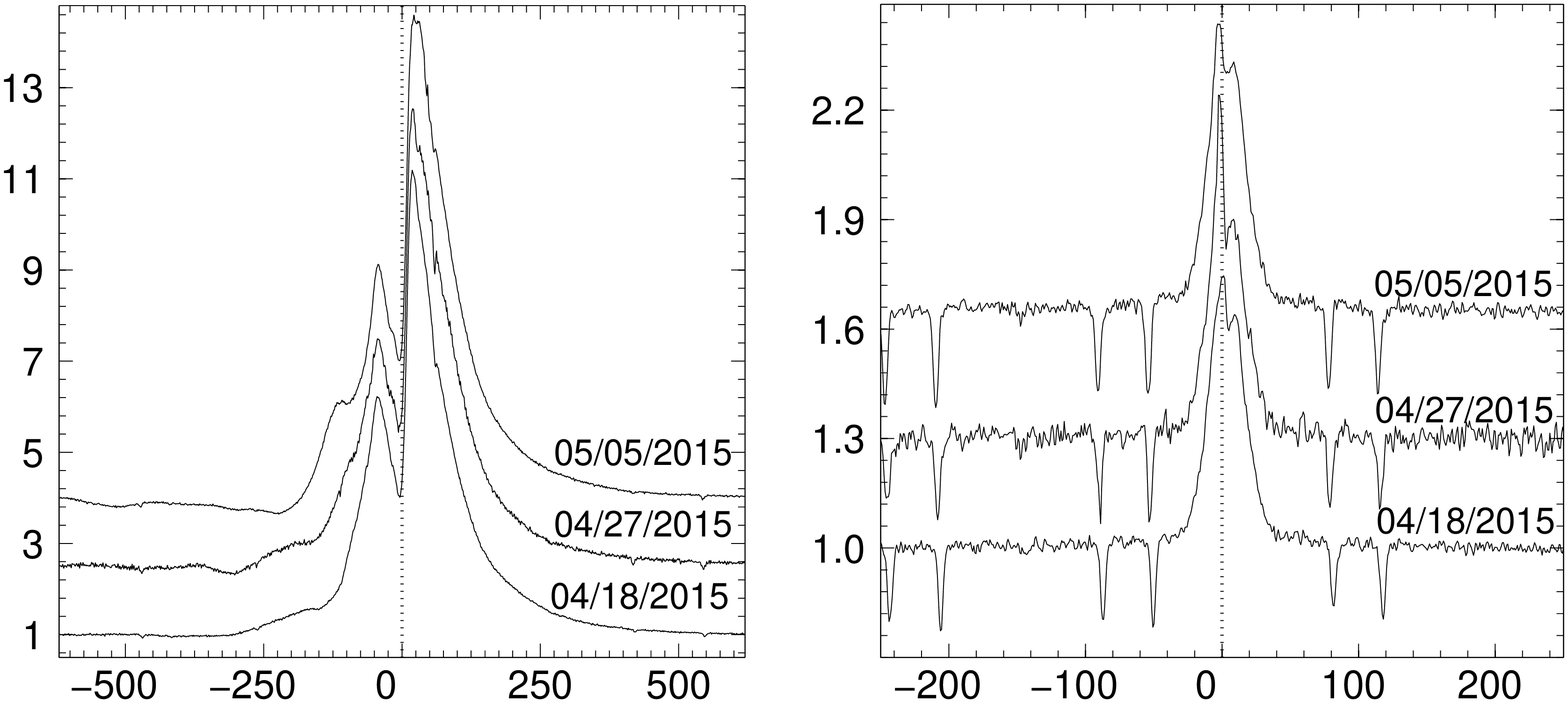}
\end{center}
\caption{{\bf Left panel:} Variations of the H$\alpha$ line in the spectrum of HD\,85567.
{\bf Right panel:} Variations of the O {\sc i} 6300 \AA\ line in the spectrum of HD\,85567. All the absorption lines shown in this panel are telluric.
Both the intensity and radial velocity are in the same units as in Fig.\,\ref{f4}.
\label{f6}}
\end{figure}

\begin{figure*}
\begin{center}
\includegraphics[width=17cm, height=11cm]{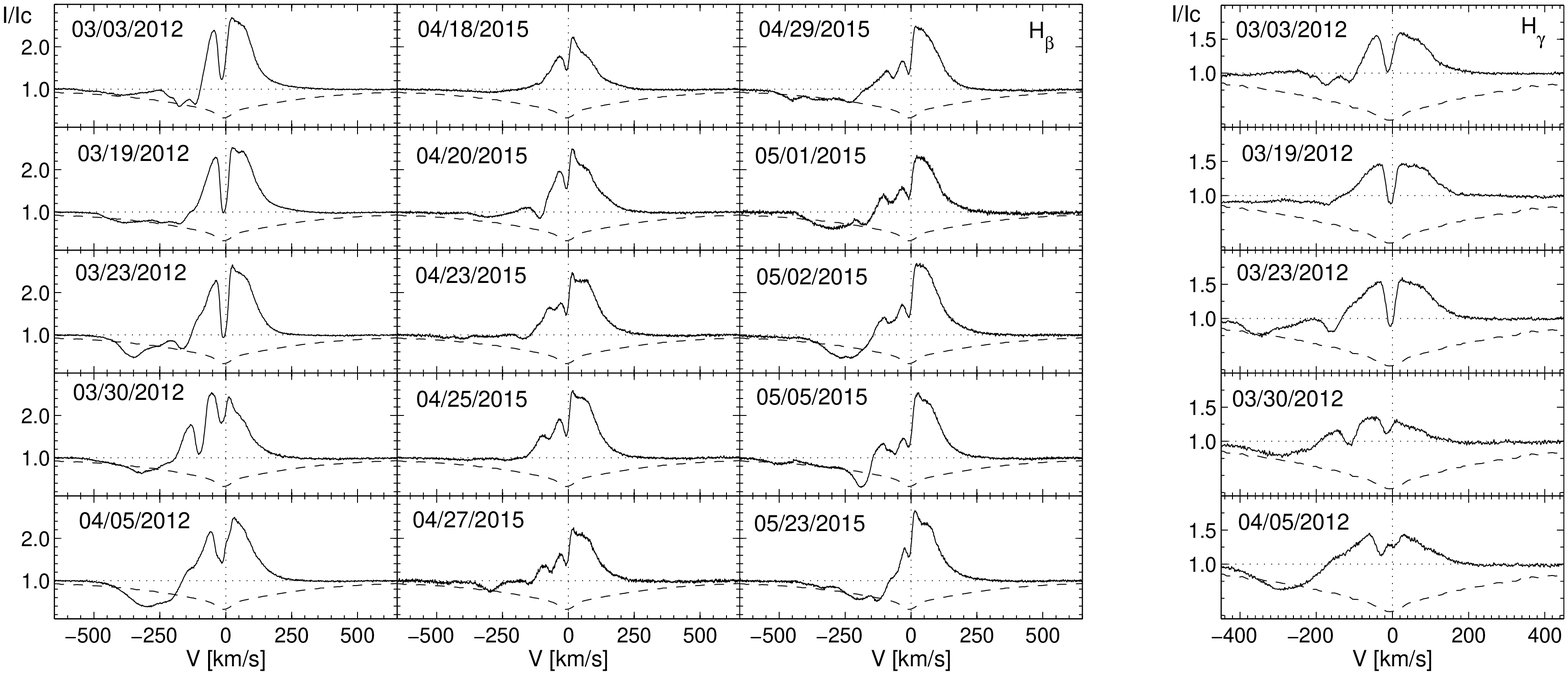}
\end{center}
\caption{ {\bf Left panel:} Variations of the H$\beta$ line in the spectrum of HD\,85567. Horizontal dotted lines show the continuum level across the line profiles, and vertical dotted lines show the systemic radial velocity. {\bf Right panel:} Variations of the H$\gamma$ line in the spectrum of HD\,85567.
Dashed lines show an expected profile of a normal star atmosphere with
T$_{\rm eff} \sim$ 15000 K, $\log g \sim$4.0, and $v \sin i \sim$30 km\,s$^{-1}$. We used the H$\beta$ and H$\gamma$ profiles of BS\,1149 for this purpose.
Both the intensity and radial velocity are in the same units as in Fig.\,\ref{f5}.
\label{f8}}
\end{figure*}

The emission-line spectrum is represented by hydrogen lines of the Balmer and Paschen series, Fe {\sc ii} lines, the [O {\sc i}] 6300 \AA\ and 6364 \AA\ lines,
the O {\sc i} 7772--7775 \AA\ triplet, and the Ca {\sc ii} triplet 8498, 8542, and 8662 \AA.
 The  Fe {\sc ii}, [O {\sc i}], and Paschen lines typically have double-peaked profiles.
The Fe {\sc ii} lines peak at I/I$_{\rm cont}\le1.2$ and show profile variations similar to those of the Balmer lines (see Fig.\,\ref{f5}, \ref{f6}, and \ref{f8}).
The Paschen lines peak at  I/I$_{\rm cont}\sim1.2$ and are less variable.
We measured EWs of the Fe {\sc ii} 5018 and 5317 \AA\ lines (EW =$-0.54\pm$0.03 \AA\ and $-0.33\pm$0.02 \AA, respectively)
as well as the [O {\sc i}] 6300, 6364 \AA\ lines (EW =$-0.45\pm0.01$ \AA\ and $-0.17\pm$0.01 \AA, respectively) and found them to show little variations.
We did a similar analysis
for the Paschen line P14, which is not distorted by other lines (e.g., Ca {\sc ii}, see Fig.\,\ref{f1}). It shows a stable weak emission component.
The RV$_{\rm blue} =-49.0\pm$3.5 km\,s$^{-1}$, RV$_{\rm red} = 39.4\pm$4.1 km\,s$^{-1}$, and EW$=-1.69\pm$0.06 \AA.

\begin{table*} [!htb]
\caption[] {RVs of some  emission lines in the spectra of HD\,85567}\label{t4}
\begin{center}
\begin{tabular}{ccccccccrc}
\hline\noalign{\smallskip}
Date  & MJD &Fe {\sc ii}$_{\rm blue}$ &Fe {\sc ii}$_{\rm red}$ &Fe {\sc ii}$_{\rm blue}$ &Fe {\sc ii}$_{\rm red}$
&Fe {\sc ii}$_{\rm blue}$ &Fe {\sc ii}$_{\rm red}$
 &[O {\sc i}] &[O {\sc i}] \\
        &           & 5018 &5018 &5169 &5169 &5317 &5317  &6364 & 6300 \\
\noalign{\smallskip}\hline\noalign{\smallskip}
03/03/2012   & 5989.642 &$-$50.5 & 63.0 & $-$49.6  & 64.8  & $-$64.4  & 58.6 &     0.3 & $-$1.2  \\
03/19/2012   & 6005.548 &$-$57.4 & 61.0 & $-$58.0  & 64.4  & $-$69.4  & 57.0 &     0.5 & $-$1.0  \\
03/23/2012   & 6009.625 & $-$59.8 & 63.4 & $-$61.5  & 72.0  & $-$67.7 & 62.6 &     0.0 & $-$1.0  \\
03/30/2012   & 6016.682 &$-$59.2 & 64.6 & $-$54.6  & 33.1  &  $-$        & $-$   & $-$0.5 & $-$1.0  \\
04/05/2012   & 6022.609 &$-$59.8 & 61.6 & $-$72.5  & 76.6  & $-$78.4  & 57.6 &     0.5 & $-$1.5  \\
04/18/2015   & 7131.546 &$-$65.2 & 69.3 & $-$58.6  & 78.9  & $-$68.8  & 64.9 & $-$0.5 & $-$1.9  \\
04/20/2015   & 7133.589 &$-$62.2 & 68.1 & $-$62.1  & 81.8  & $-$65.5  & 65.5 & $-$0.5 & $-$1.9  \\
04/23/2015   & 7136.553 &$-$67.0 & 59.8 & $-$59.2  & 77.8  & $-$65.5  & 59.2 & $-$0.5 & $-$1.9  \\
04/25/2015   & 7138.621 &$-$58.6 & 68.7 & $-$55.1  & 76.0  & $-$60.9  & 64.9 & $-$0.5 & $-$1.9  \\
04/27/2015   & 7140.610 &$-$56.2 & 72.9 & $-$56.3  & 72.5  & $-$71.7  & 60.4 & $-$0.5 & $-$2.9  \\
04/29/2015   & 7142.619 &$-$65.2 & 72.9 & $-$66.7  & 84.2  & $-$67.1  & 66.0 & $-$0.5 & $-$1.9  \\
05/01/2015   & 7144.598 &$-$60.4 & 68.1 & $-$48.8  & 77.2  & $-$69.4  & 67.7 & $-$1.4 & $-$2.4  \\
05/02/2015   & 7145.577 &$-$64.6 & 67.0 & $-$64.4  & 82.4  & $-$71.1  & 66.6 & $-$0.5 & $-$2.4  \\
05/05/2015   & 7148.612 &$-$63.4 & 74.7 & $-$60.9  & 81.8  & $-$69.4  & 64.9 & $-$0.5 & $-$1.9  \\
05/23/2015   & 7166.499 &$-$66.4 & 72.3 & $-$61.5  & 76.0  & $-$71.7  & 70.0 & $-$0.5 & $-$1.9  \\
\noalign{\smallskip}\hline
\end{tabular}
\end{center}
\begin{list}{}
\item Column information: (1) -- observing date, (2) -- JD$-$2450000, (3)--(8) -- heliocentric RVs of some Fe {\sc ii} lines in km\,s$^{-1}$,\\
(9)--(10) heliocentric RVs of the [O {\sc i}] line in km\,s$^{-1}$.
\item The 5169 \AA\ line exhibits an additional blue-shifted emission peak at RV = $-$132 km\,s$^{-1}$ in the spectrum\\ taken on 03/30/2012   (see Fig.\,\ref{f5} and Sect.\,\ref{discussion}).
\item Rest wavelengths for the [O {\sc i}] lines, 6300.33 and 6363.77 \AA, were taken from the Atomic Line List by P.~van Hoof available from
http://www.pa.uky.edu/$\sim$peter/newpage/
\end{list}
\end{table*}

\begin{figure} [!htb]
\begin{center}
\includegraphics[width=9.0cm, height=5.0cm]{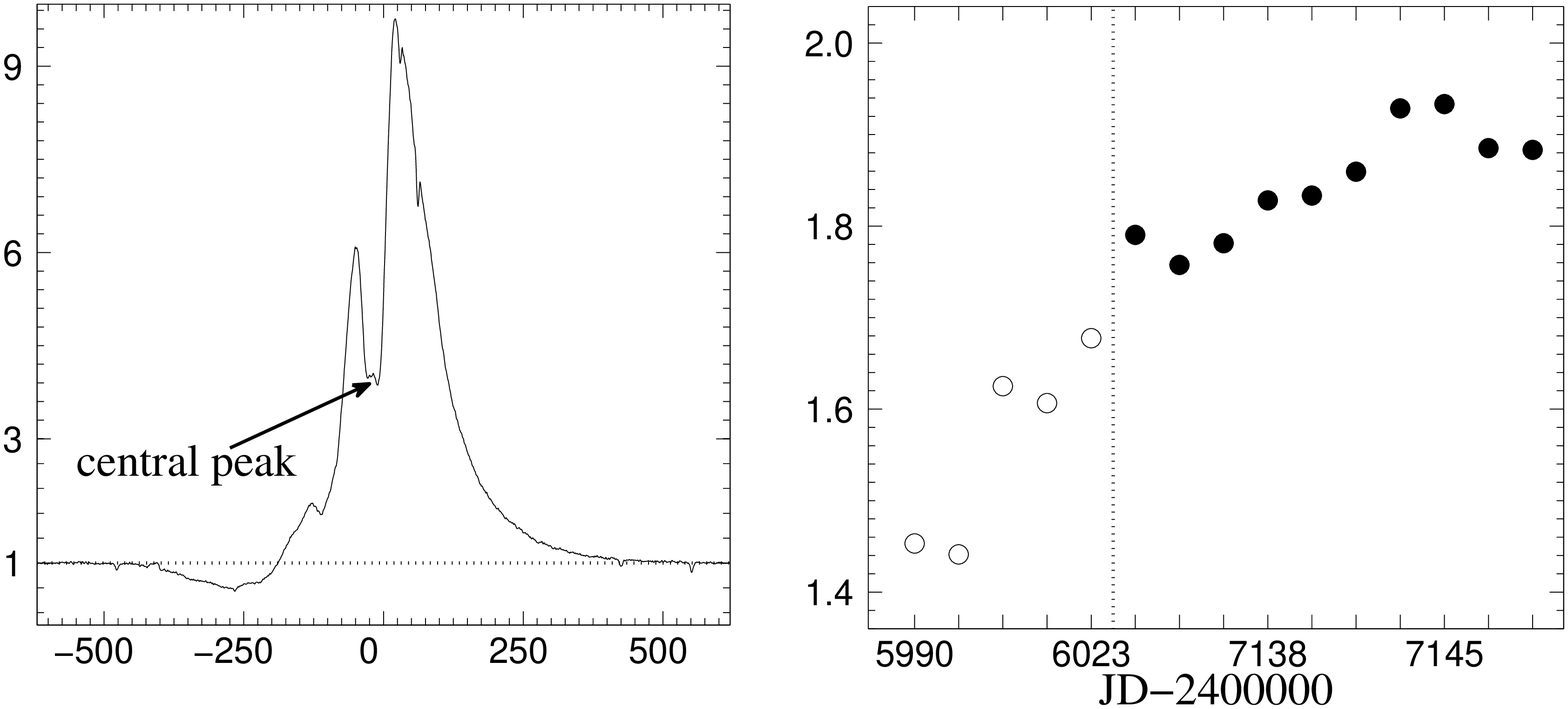}
\end{center}
\caption{{\bf Left panel:} The H$\alpha$ line profile in the spectrum of HD\,85567 when a central peak is observed. Both the intensity and radial velocity are in the same units as in Fig.\,\ref{f5}.
{\bf Right panel:} Temporal evolution of the intensity ratio of the red to the blue peak ($I_{\rm red}/I_{\rm blue}$) of
the H$\alpha$ line in the spectrum of HD\,85567. The data from 2012 are shown by open circles, and the data from 2015 are shown by filled circles.
\label{f7}}
\end{figure}

The H$\alpha$ and H$\beta$ lines show strong emission components,  which are dominated by a double-peaked profile. Unlike the other lines with nearly equal peak strengths (including H$\gamma$), these two lines exhibit a stronger redshifted emission peak.
However, sometimes a third peak appears on their blue side and moves toward the double-peaked structure (see Fig.\,\ref{f6} for the H$\alpha$ line).
Another feature of the Balmer line profiles is a blue-shifted absorption component which also varies in shape and position. Additionally, the Balmer lines occasionally exhibit more complicated profiles, such as a central peak in the H$\alpha$ line shown in Fig.\,\ref{f7}.  The H$\beta$ and H$\gamma$ line profiles are highly variable on a time scale of a few days (Fig.\,\ref{f8}). The H$\alpha$ line profile looks less variable (Fig.\,\ref{f6}), because the moving features are less noticeable against its stronger emission components.

RVs of some Fe {\sc ii} and [O {\sc i}] lines are shown in Table\,\ref{t4}.
Parameters of the H$\alpha$, H$\beta$, are H$\gamma$ lines are presented in Tables\,\ref{t5}, \ref{t6}, and \ref{t7}, respectively.

\begin{table}[!h]
\caption[] {Parameters of the H$\alpha$ line in the spectra of HD\,85567}\label{t5}
\begin{center}
\begin{tabular}{ccrcrc}
\hline\noalign{\smallskip}
Date       &RV$_{\rm blue}$ & RV$_{\rm red}$ &I$_{\rm blue}$ &I$_{\rm red}$  &$EW$\\
\noalign{\smallskip}\hline\noalign{\smallskip}
03/03/2012    & $-$50.7  & 23.3   & 6.4 & 9.3   & $-$29.9 \\
03/19/2012    & $-$43.9  & 28.3   & 6.8 & 9.8   & $-$33.1 \\
03/23/2012    & $-$41.1  & 26.5   & 6.4 & 10.4  & $-$32.8 \\
03/30/2012    & $-$49.4  & 21.5   & 6.1 & 9.8   & $-$29.0 \\
04/05/2012    & $-$45.3  & 24.7   & 6.2 & 10.4  & $-$32.0 \\
04/18/2015    & $-$43.0  & 19.7   & 6.2 & 11.1  & $-$31.5 \\
04/20/2015    & $-$43.4  & 18.7   & 6.6 & 11.6  & $-$34.3 \\
04/23/2015    & $-$43.4  & 19.2   & 6.4 & 11.4  & $-$35.0 \\
04/25/2015    & $-$43.0  & 19.2   & 6.4 & 11.7  & $-$36.3 \\
04/27/2015    & $-$43.0  & 19.7   & 6.0 & 11.0  & $-$34.0 \\
04/29/2015    & $-$43.0  & 20.1   & 6.4 & 11.9  & $-$31.2 \\
05/01/2015    & $-$42.1  & 19.7   & 5.6 & 10.8  & $-$31.9 \\
05/02/2015    & $-$42.5  & 20.1   & 6.0 & 11.6  & $-$35.3 \\
05/05/2015    & $-$42.5  & 21.5   & 6.1 & 11.5  & $-$35.2 \\
05/23/2015    & $-$39.3  & 20.1   & 6.0 & 11.3  & $-$31.7 \\
\noalign{\smallskip}\hline
\end{tabular}
\end{center}
\begin{list}{}
\item The observing date is shown in Col. 1. The heliocentric RVs or the blue (RV$_{\rm blue}$), red (RV$_{\rm red}$),
 emission peak are listed in Cols. 2 and 3  respectively; their intensities in continuum
units are listed in Cols. 4 and 5; and $EW$s of the line emission component are given in \AA\ in Col. 6.
\end{list}
\end{table}

\begin{table}[!h]
\caption[] {Parameters of the H$\beta$ line in the spectra of HD\,85567}\label{t6}
\begin{center}
\begin{tabular}{ccrrcccc}
\hline\noalign{\smallskip}
Date       &RV$_{\rm blue}$ & RV$_{\rm red}$ &RV$_{\rm add}$ &I$_{\rm blue}$ &I$_{\rm red}$ &I$_{\rm add}$ &$EW$\\
\noalign{\smallskip}\hline\noalign{\smallskip}
03/03/2012    &$-$46.3 &25.3&         & 2.3  & 2.6  &      &$-$4.4 \\
03/19/2012    &$-$40.1 &29.6&         & 2.3  & 2.5  &      &$-$4.4 \\
03/23/2012    &$-$37.6 &27.2&         & 2.3  & 2.6  &      &$-$4.2 \\
03/30/2012    &$-$53.1 &13.0&$-$133.3 & 2.5  & 2.4  & 1.8  &$-$4.5\\
04/05/2012    &$-$58.0 &33.3&         & 2.1  & 2.5  &      &$-$4.2\\
04/18/2015    &$-$33.3 &17.9&         & 1.7  & 2.2  &      &$-$2.8\\
04/20/2015    &$-$33.3 &16.7&         & 2.0  & 2.5  &      &$-$3.4\\
04/23/2015    &$-$28.4 &15.4&$-$71.0  & 1.7  & 2.4  & 1.6  &$-$3.8\\
04/25/2015    &$-$33.9 &16.0&$-$96.9  & 1.9  & 2.6  & 1.5  &$-$4.1\\
04/27/2015    &$-$29.0 &17.9&$-$92.0  & 1.7  & 2.5  & 1.4  &$-$2.9\\
04/29/2015    &$-$32.1 &19.1&$-$91.3  & 1.7  & 2.5  & 1.4  &$-$3.6\\
05/01/2015    &$-$33.9 &29.0&$-$103.7 & 1.6  & 2.3  & 1.4  &$-$3.1\\
05/02/2015    &$-$32.1 &29.0&$-$102.4 & 1.7  & 2.7  & 1.4  &$-$4.1\\
05/05/2015    &$-$29.0 &26.5&$-$108.6 & 1.4  & 2.5  & 1.34 &$-$3.3\\
05/23/2015    &$-$24.1 &16.7&         & 1.8  & 2.6  &      &$-$3.4\\
\noalign{\smallskip}\hline
\end{tabular}
\end{center}
\begin{list}{}
\item The observing date is shown in Col. 1. The heliocentric RVs or the blue (RV$_{\rm blue}$), red (RV$_{\rm red}$),
and an additional, blue-shifted (RV$_{\rm add}$) emission peak are listed in Cols. 2--4, respectively; their intensities in continuum
units are listed in Cols. 5--7; and the $EW$'s of the line emission component are given in \AA\ in Col. 8.
\end{list}
\end{table}

\begin{table}[!h]
\caption[] {Parameters of the H$\gamma$ line in the spectra of HD\,85567}\label{t7}
\begin{center}
\begin{tabular}{ccrcccc}
\hline\noalign{\smallskip}
Date       &RV$_{\rm blue}$ & RV$_{\rm red}$ &RV$_{\rm add}$ &I$_{\rm blue}$ &I$_{\rm red}$ &I$_{\rm add}$\\
\noalign{\smallskip}\hline\noalign{\smallskip}
03/03/2012    & $-$35.1  & 32.5  &          & 1.5 & 1.6 &      \\
03/19/2012    & $-$33.2  & 32.5  &          & 1.5 & 1.5 &      \\
03/23/2012    & $-$33.2  & 29.7  &          & 1.5 & 1.6 &      \\
03/30/2012    & $-$51.1  & 15.9  & $-$145.1 & 1.4 & 1.3 & 1.1  \\
04/05/2012    & $-$58.8  & 33.1  &          & 1.4 & 1.4 &      \\
\noalign{\smallskip}\hline
\end{tabular}
\end{center}
\begin{list}{}
\item The column information is the same as in Table\,\ref{t5}, except for the EW which was not measured because of the line complicated profile  (see Fig.\,\ref{f8}).
\end{list}
\end{table}

\subsection{Brightness variations}\label{brightness}

\begin{figure}
\begin{center}
\includegraphics[width=9cm, height=7cm]{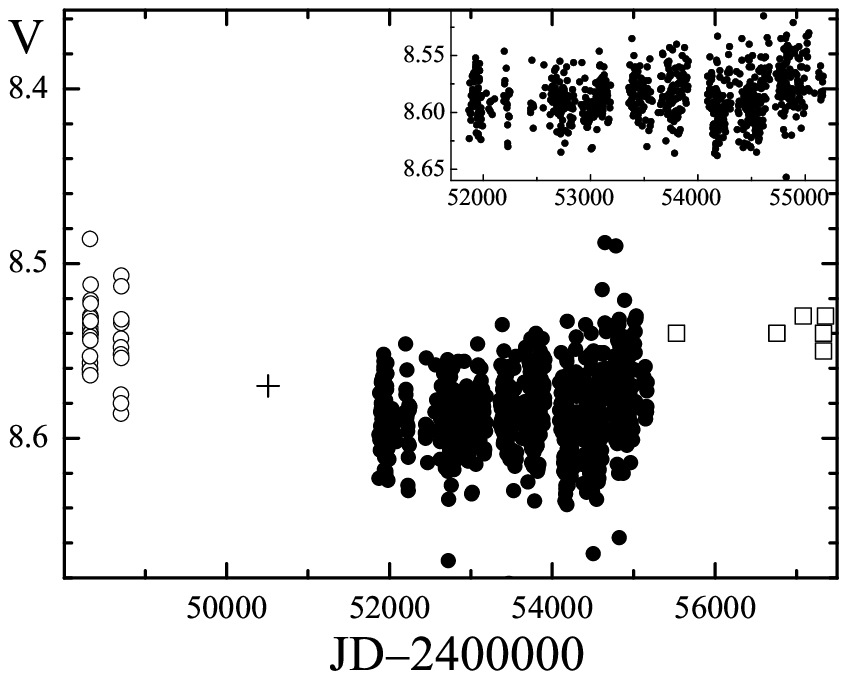}
\end{center}
\caption{{The $V$--band light curve of HD\,85567. Open circles -- data from \citet{2001A&A...380..609D}, the plus -- data from \citet{2001A&A...371..600M},
filled circles -- data from the ASAS--3 survey \citep{2002AcA....52..397P}, and  open squares -- our PROMPT data. The inset shows ASAS--3 data only. Typical brightness measurement uncertainty is 0.02 mag.}
\label{f10}}
\end{figure}

Photometric observations of HD\,85567 have been published in several papers.
\citet{1977A&AS...27..215K}, \citet{1983ApJS...51..321S}, and \citet{2001A&A...371..600M} each presented two
measurements of the optical brightness obtained in 1973--1975 ($V$ = 8.55 mag), 1968--1977 ($V$ = 8.58 mag), and 1997, respectively. \citet{2001A&A...380..609D}
obtained 33 data points during two week-long periods in 1991 and 1992. The star was observed in the ASAS--3 survey in 2004--2010 \citep{2002AcA....52..397P}.  Our PROMPT data include six measurements of the optical brightness obtained in 2010--2015 ($V$ = 8.54$\pm$0.01 mag).
The $V$--band light curve in the last 25 years (see Fig.\,\ref{f10}) exhibits neither significant nor regular variations.

\begin{figure}
\begin{tabular}{cc}
\hspace*{-0.7cm}\resizebox{5.2cm}{!}{\includegraphics{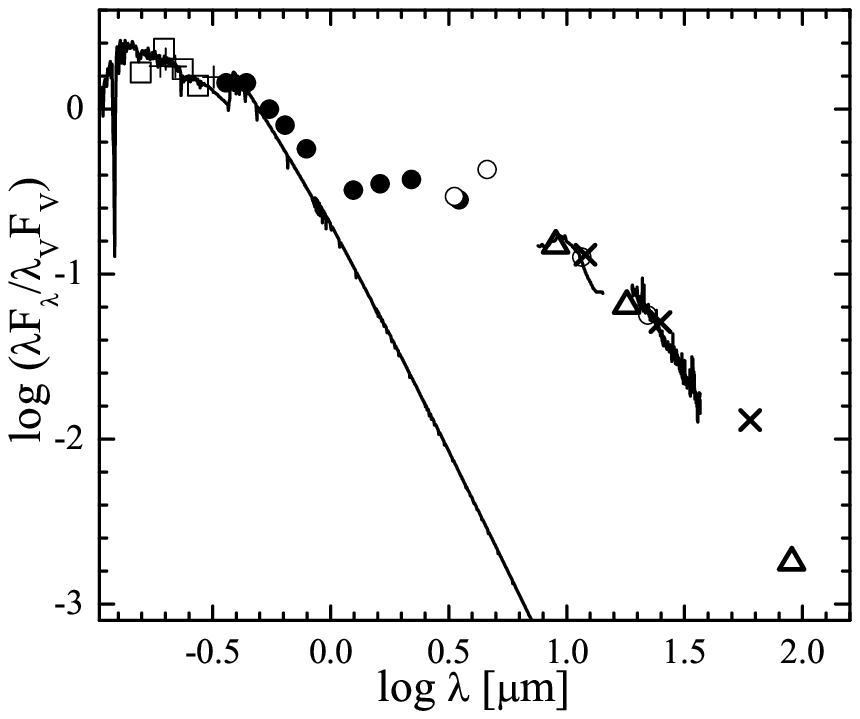}}&
\hspace*{-0.8cm}\resizebox{5cm}{!}{\includegraphics{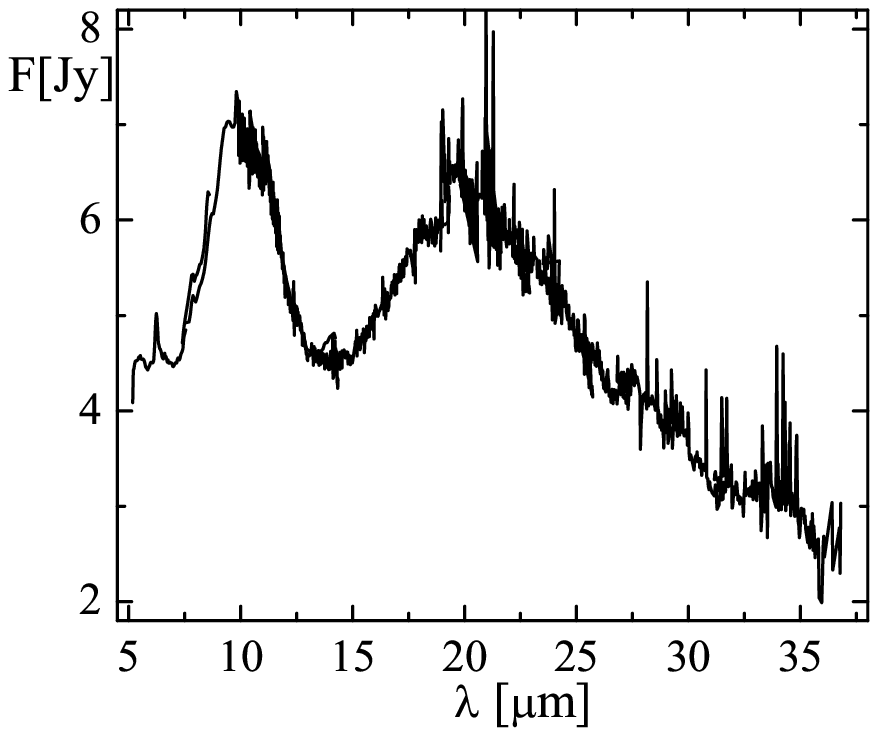}}\\
\end{tabular}
\caption{{\bf Left panel:} Spectral energy distribution of HD\,85567 from ground-based and space data. Symbols: pluses -- TD1 satellite
photometric data \citep{1978csuf.book.....T}, open squares -- fluxes measured from the IUE spectrum LWP19729LL, filled circles --
ground-based photometry \citep{2001A&A...371..600M}, open circles -- WISE data, triangles -- AKARI data, crosses -- IRAS data, and
short solid lines over the IR data points -- parts of the Spitzer Space Observatory spectrum shown in the right panel. The interstellar
extinction (A$_V$ = 0.50 mag, see text) was removed using a wavelength dependence from \citet{1979ARA&A..17...73S}.
The solid lines over the UV and visual region data represent a model atmosphere for T$_{\rm eff}$=15000 K and $\log$ g = 4.0 \citep{1998HiA....11..646K}.
{\bf Right panel:} The Spitzer Space Observatory spectrum of HD\,85567 first published by \citet{2010ApJ...721..431J}.
\label{f9}}
\end{figure}

\subsection{Interstellar extinction law in the direction toward HD\,85567}\label{extinction}

\begin{figure}
\includegraphics[width=8.5cm, height=5.0cm]{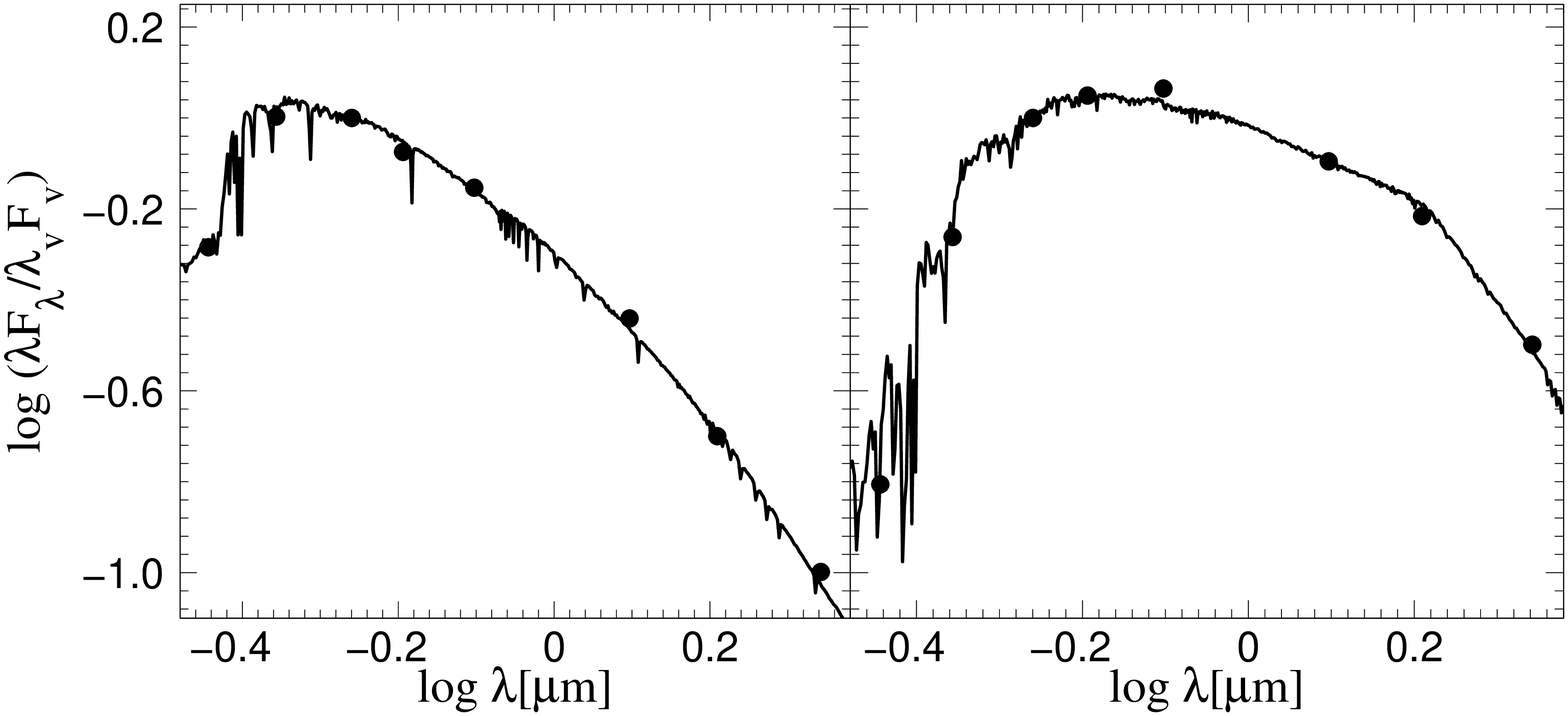}
\caption{{\bf Left panel:} A de-reddened SED for star No. 15 from Table\,\ref{t2}. The photometric $UBV(RI_{\rm c})JHK$ data are shown by filled circles.
The solid line represents the intrinsic SED for $T_{\rm eff}=$7000 K and $\log g=$4.0 from \citet{1998HiA....11..646K}. For the star's A$_V$ see Table\,\ref{t2}.
{\bf Right panel:} A de-reddened SED for star No. 2 from Table \ref{t2}. The photometric $UBV(RI_{\rm c})HK$ data are shown by filled circles.
The solid line represents the intrinsic SED for $T_{\rm eff}=$5000 K and $\log g=$3.0 from \citet{1998HiA....11..646K}.
\label{f11}}
\end{figure}

\begin{figure}
\begin{center}
\includegraphics[width=8cm, height=6cm]{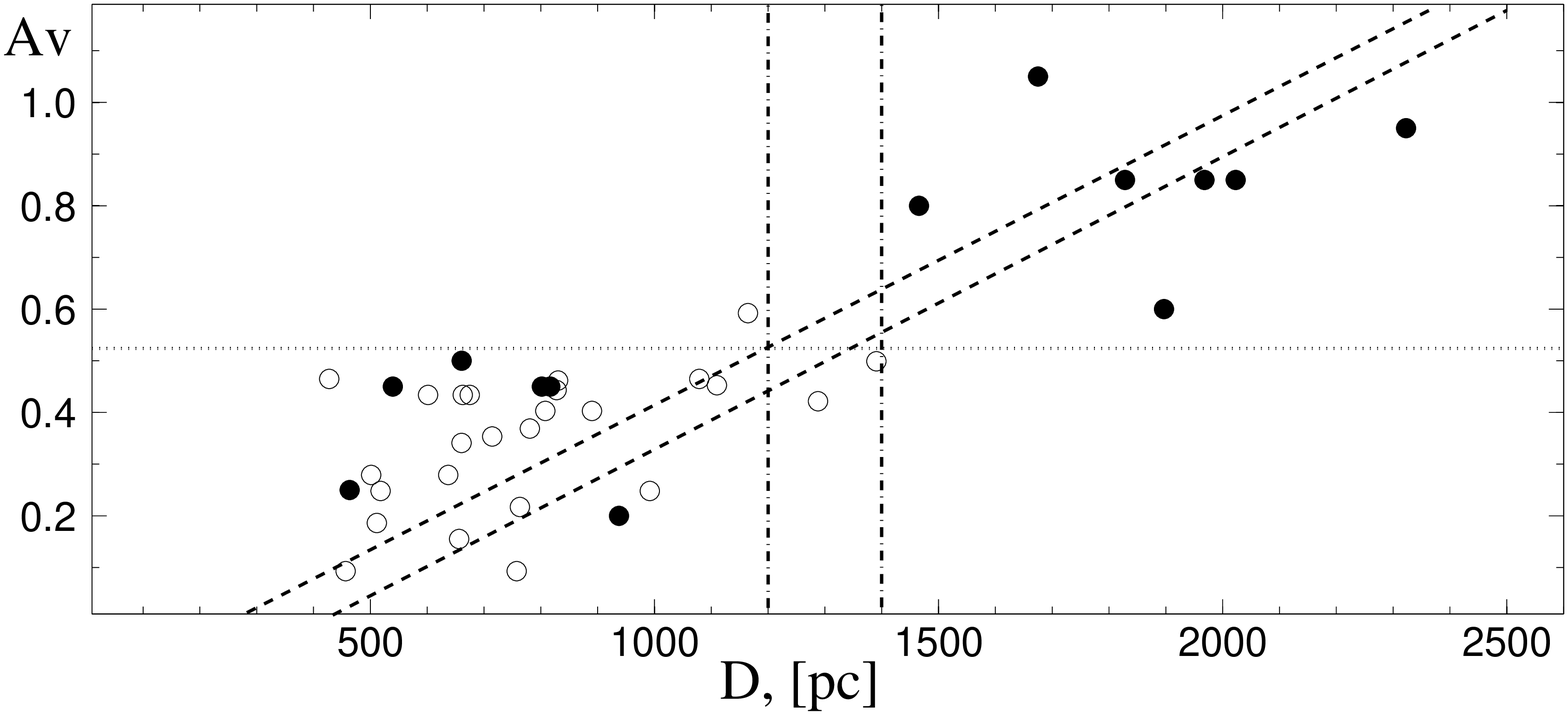}
\end{center}
\caption{{A distance dependence of the interstellar extinction toward HD\,85567. Data for the stars marked in Fig.\,\ref{f2} are shown by filled circles and for stars
from a 60$\arcmin$ region around the object are shown by open circles. The diagonal dashed lines show a 1--$\sigma$ deviation from the average interstellar
extinction law in the area. Stars with more that a 3--$\sigma$ deviation from the average relationship were removed from the fit. The vertical dashed lines show
a range of the distance uncertainty derived for HD\,85567 from the extinction law using the value of A$_V$ found from the $EW$ of the 5780 \AA\ DIB shown by the horizontal dashed line (see Sect.\,\ref{absorptions}).}
\label{f12}}
\end{figure}

To determine the distance toward HD\,85567, we constructed a relationship of the interstellar extinction with distance from the Sun in its direction (see Fig.\,\ref{f12}). To do that, we collected published photometric data and MK types for hot stars in a region of 60$\arcmin$ from the star position. Additionally, we measured brightness of some stars in a $10\arcmin \times 10\arcmin$ field around HD\,85567 (see Fig.\,\ref{f2}) using our $UBV(RI)_{\rm c}$ data in the Johnson-Cousins photometric system obtained at PROMPT telescopes. We also added 2MASS $JHK$ data \citep{2003yCat.2246....0C} to our optical photometry. We determined $T_{\rm eff}$ of these stars by comparison with model SEDs from \citet{1998HiA....11..646K}. The observed spectral energy distributions (SEDs) were de-reddened using a standard Galactic interstellar extinction law \citep{1979ARA&A..17...73S}. Two examples of the SED fitting are shown in Fig.\,\ref{f11}. Distances to the stars were determined using a MK type luminosity calibration from \citet{1981Ap&SS..80..353S}. The distance to HD\,85567 was determined from a linear fit to the data shown in Fig.\,\ref{f12}  using the value of the interstellar extinction from the 5780 \AA\ DIB EW (see Sect.\,\ref{absorptions}). This result is in agreement with the RV of the absorption lines due to Galactic rotation \citep[see discussion in][]{2001A&A...371..600M}.

\section{Discussion}\label{discussion}

\subsection{Fundamental parameters}\label{parameters}

It follows from analysis of the literature data (see Table \ref{t1}) that fundamental parameters of HD\,85567 are not well constrained. Most papers that mention HD\,85567 usually use the values derived in previous studies. However, we suggest that choices for some of the parameters have not been sufficiently justified.
For example, \citet{2001A&A...371..600M} favored a spectral type B2 but mentioned a discrepancy between the observed and theoretical absorption line profiles.
Fundamental parameters are a very important factor for understanding the nature and evolutionary status of the star. Therefore, we attempted a new derivation using our
more extended data set that includes multicolor photometry of HD\,85567 and stars in its vicinity as well as our new high-resolution spectra.

Based on the results derived in Sect.\,\ref{extinction}, the most probable distance to HD\,85567 is 1.3$\pm$0.1 kpc (see Fig.\,\ref{f12}). It is consistent with a negative result
on the object's parallax derived by the {\it HIPPARCOS} mission \citep{1997ESASP1200.....E,2007A&A...474..653V}. The value of $T_{\rm eff}=$15000$\pm$500 K corresponds to a spectral type B4/B5 (Sect.\ref{absorptions}). This leads to a luminosity of $\log$ L/L$_\odot$ = 3.4$\pm$0.1 calculated from the average observed visual magnitude $V$ = 8.55$\pm$0.05 mag, interstellar extinction A$_{V} = 0.50\pm0.02$ mag, and a bolometric correction for the adopted $T_{\rm eff}$, BC$(V) = -1.2$ mag \citep{1997IAUS..189P..50M}.

This luminosity estimate does not take into account a possible contribution from the circumstellar disk, which manifests itself by the double-peaked emission-line profiles and
strong IR excess. Since the absorption lines are not significantly weaker than those of normal stars with similar T$_{\rm eff}$ and the disk tilt angle \citep[$i=53\degr$,][]{2014A&A...569A..25V} is not very large, we adopt the disk contribution to the visual brightness of $\le$ 20\%. This estimate is consistent with the observed color-index
$B-V \sim 0.1$ mag which is $\sim 10$\% larger than the one expected from the interstellar extinction derived from the DIB strength.

With this correction, the luminosity of HD\,85567 becomes $\log$ L/L$_\odot$ = 3.3$\pm$0.2. It is in agreement with that currently adopted for a single $\sim$6 M$_\odot$ star near the end of the main-sequence evolutionary stage \citep[][see Fig.\,\ref{f13}]{2012A&A...537A.146E}. Finally, combining the disk tilt angle and
 $v \sin\,i = 31\pm3$ km\,s$^{-1}$ (see Sect.\,\ref{absorptions}), a full rotational velocity of the star is 37$\pm$3 km\,s$^{-1}$.
The fundamental parameters of HD\,85567 are listed in  Table \ref{f1}.

\subsection{Evolutionary state of the HD\,85567}\label{Evolutionary}

As mentioned in Sect.\ref{intro}, there is a disagreement about the evolutionary state of HD\,85567. Here we review several possibilities to re-analyze this subject.

\subsubsection{Single star}\label{hbe}

If the star is currently single, then it could have been formed single or become a binary merger product. First, we review a suggestion that HD\,85567 might be a HAeBe star initially proposed by \citet{1994A&AS..104..315T} and carried on in a number of later papers \citep[e.g.,][]{2012A&A...538A.101V,2013A&A...558A.116W}. This requires a comparison of the observed properties of HD\,85567 with those of {\it bona fide} HAeBe stars. \citet{2012A&A...538A.101V} presented SEDs of a group of HAeBe stars including HD\,85567.

The main distinguishing feature of our star is a lack of the far-IR excess, which is represented by a sharply declining flux at wavelengths $\lambda > 10 \mu$m (see Fig.\,\ref{f9}). This feature is due to a lack of cold/distant dust in the circumstellar environments of HD\,85567. \citet{2012A&A...538A.101V} explain this feature by photo-evaporation of the circumstellar disk. \citet{2013A&A...558A.116W} also argued that the absence of the far-IR excess suggested that HD\,85567 is photo-evaporating its disc from outside. However the region of the sky around the star (see Fig.\,\ref{f2}) contains no hot stars which could provide high-energy radiation to destroy the dust near it.

\begin{figure}
\begin{center}
\includegraphics[width=9cm, height=5cm]{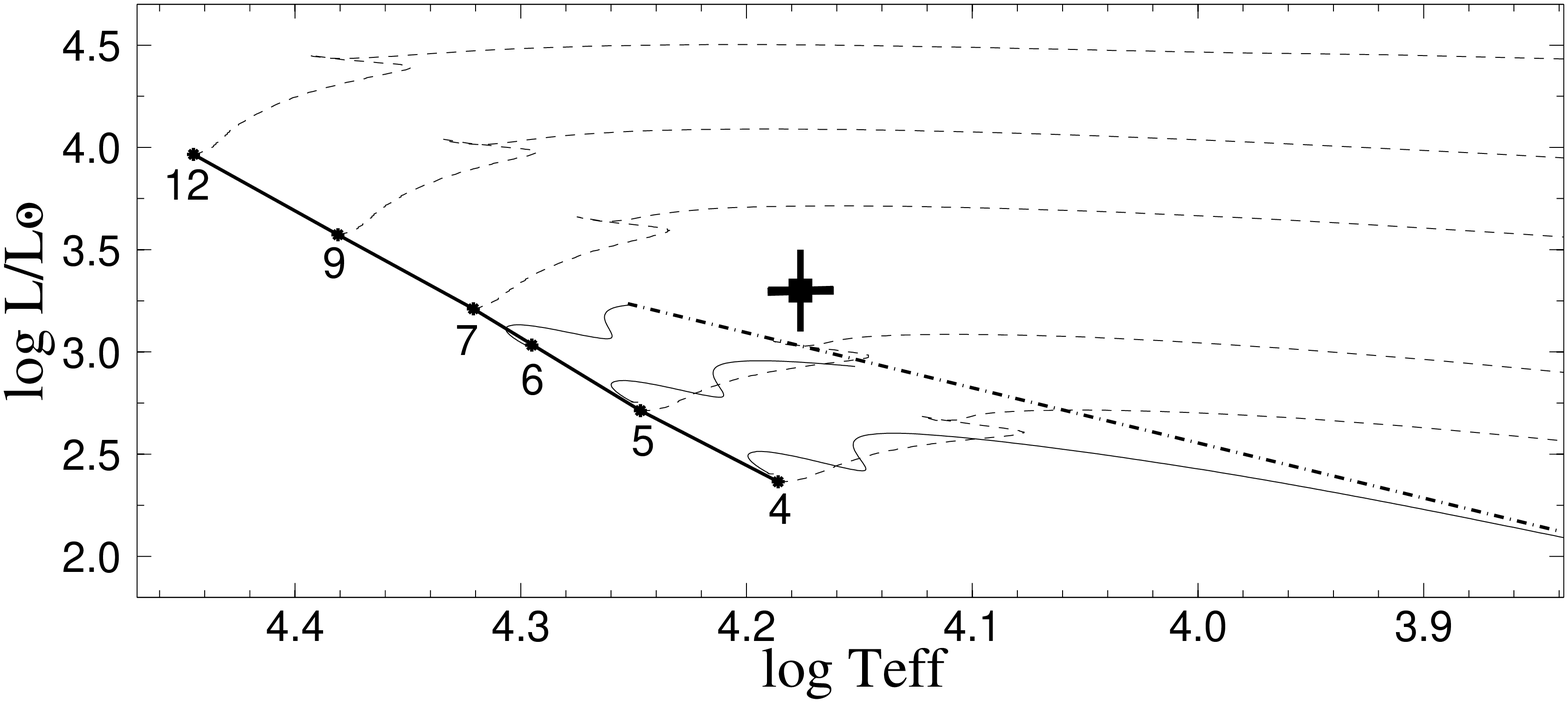}
\end{center}
\caption{{The Hertzsprung–--Russell diagram showing evolutionary tracks of pre-main-sequence stars  (4, 5, 6 M$_\odot$) from \citet{2011yCat..35330109T} (thin solid lines) and rotating single stars  (4, 5, 7, 9, 12 M$_\odot$) from \citet{2012A&A...537A.146E} (dashed lines). The thick solid line represents the Zero-Age Main-Sequence (ZAMS), the dash-dotted line represents the birth-line for pre-main-sequence stars, ZAMS masses are indicated by numbers in solar units near corresponding tracks.
The position of HD\,85567 is indicated by a filled square using the fundamental parameters from  Table\,\ref{t1}.}
\label{f13}}
\end{figure}

Also from a theoretical point of view, \citet{2015ApJ...804...29G} show that the inner disk is dispersed before the outer disk. Moreover, \citet{2009ApJ...690.1539G} show that photo-evaporation from the central star occurs from the disk surface (gas coupled with small dusty particles), where optical depth due to dust is of the order of unity and where FUV photons from the star begin to get attenuated, i.e. where A$_V$ = 0.1--"0.5 mag. This indicates that the dust should still remain in the outer disk. The photo-evaporation theory also suggests that the outer dusty disk may quickly evaporate at the latest stages of pre-main-sequence evolution. Nevertheless, the fundamental parameters of HD\,85567 determined here are inconsistent with a those of HAeBe stars approaching the main-sequence (see Fig.\,\ref{f13}). Additionally, a B5--type star may not be able to provide a high FUV flux sufficient for photo-evaporation to work efficiently.

\begin{table}[!h]
\caption[] {Comparison of fundamental parameters of HD\,85567 under an assumption of a pre-main-sequence star}\label{t9}
\begin{center}
\begin{tabular}{cccc}
\hline\noalign{\smallskip}
Mass        & $\log$ L   & R$_V$=3.1 & R$_V$=5\\
M$_\odot$ & L$_\odot$    &  D, pc   &  D, pc\\
\noalign{\smallskip}\hline\noalign{\smallskip}
3.6    & 2.31  & 342   & 325  \\
4.0    & 2.60  & 473   & 449  \\
5.0    & 2.92  & 690   & 655  \\
6.0    & 3.16  & 909   & 862  \\
7.0    & 3.36  & 1139  & 1080 \\
\noalign{\smallskip}\hline
\end{tabular}
\end{center}
\begin{list}{}
\item .
\end{list}
\end{table}

Adopting that HD\,85567 is a single HAeBe star, we determined the distance toward it assuming  the luminosity found at the points of intersection with pre-main-sequence evolutionary tracks for different masses \citep{2011yCat..35330109T} and T$_{\rm eff}$ determined by us (see Sect.\,\ref{absorptions} and  Table\,\ref{t1}).
The results of these calculations are presented in Table\,\ref{t9}.  As shown in Fig. \ref{f13}, HD\,85567 gets further away from the ZAMS as the initial mass increases. This indicates that HD\,85567 is not a HAeBe star for masses of  M $\ge$\,6M$_\odot$, because such a pre-main-sequence star should still reside within its parental cloud and therefore is invisible at this stage.  A similar approach was adopted by \citet{2015MNRAS.453..976F}, who determined an A$_V$ of 0.89$\pm$0.03 mag from the object's color-excess $E(B-V)$ based on the observed and intrinsic color-index $B-V$ (for T$_{\rm eff} = 13000\pm500$ K they found from fitting a low-resolution optical
spectrum of HD\,85567 to model spectra of normal stars) and ignoring any circumstellar contribution to the optical continuum. These authors found a mass of M = $6^{+2.7}_{-1.8}$ M$_\odot$, which puts the star very close to the birthline for this mass range. However, one can see no nebulosity or a star-forming region in the vicinity of HD\,85567 (e.g., see Fig.\,\ref{f2}).

There remains a possibility that the star has a mass of 4 M$_\odot$ and is about to start its main-sequence life. However it does not agree with the observational data (the distance derived from the interstellar extinction, see Sect.\ref{parameters}, and the absence of an obvious star forming region around the star). Therefore, we confirm an earlier suggestion by \citet{2001A&A...371..600M} that HD\,85567 cannot be a HAeBe star.

The possibility that the star is currently at the main-sequence evolutionary stage is unlikely because its circumstellar matter then should have been produced by its own
wind, which cannot explain the observed emission-line strengths.
The theory of stellar evolution does not predict such a strong mass loss from a single, not rapidly rotating main-sequence star \citep[e.g.,][see also a discussion of this subject in \citet{m07}]{2014A&A...564A..70K}.
The possibility of a merger is more realistic but not easily
verifiable. A merger is capable of producing large amounts of circumstellar matter, although there is no evidence for the merging process. It would have accompanied by a strong brightness increase, as it was observed in the binary merger of V\,1309 Sgr \citep{2011A&A...528A.114T}. Since the object's light curve does not show any significant outbursts (see Fig.\,\ref{f10}), such an event might have occurred over 40 years ago. The most important argument against this hypothesis is that a merger would increase the rotation rate of its product, while we observe a low rotation velocity (see Sect.\,\ref{parameters}).

\subsubsection{Binary system}\label{binary}

The second scenario assumes that the observed IR-excess arises as a consequence of a non-conservative mass-transfer in a close binary system. The circumstellar dust should exist in the circumbinary area, as the components in such systems are too close together to allow its existence around either star. \citet{m07a} proposed this scenario to explain the presence of dust in FS\,CMa type objects. Evolution of intermediate-mass binary systems has been studied theoretically in several papers \citep[e.g.,][]{2008A&A...487.1129V,2015A&A...577A..55D}.

We did not find signs of a secondary component in the spectrum of HD\,85567. However, this is not an ultimate proof that such a component does not exist. They are typically much ($\Delta V > $ 2 mag) fainter than the primary B--type star in many classical Be stars and some FS\,CMa type objects. For example,
\citet{2015ApJ...809..129M} found a faint G8--type secondary the FS\,CMa type object MWC\,728 that manifests itself by very weak ($<$5\% of the continuum
intensity) absorption lines and is 6 times fainter in the $V$--band than the B5--type primary.

We can put some constraints on the mass of a hypothetical secondary in HD\,85567 by evaluating the system mass function. Assuming that the observed variation of the individual absorption line RVs (1.3 km\,s$^{-1}$, see Table\,\ref{t3}) is due to orbital motion and the orbital period is not very large \citep[typically 1--2 months following the results for binary systems that undergo a strong mass-transfer,][]{2008A&A...487.1129V}, the mass function turns out to be $\le10^{-5}$ M$_\odot$. Fourier analysis of the ASAS--3 data set (see Fig.\,\ref{f10}) shows the only significant period at 180 days. It is most likely due to availability of the object in the sky. However, using this period in calculating the mass function changes the above value insignificantly.

The latter result implies the following possibilities: (1) the secondary companion is not present (a merger has occurred which is unlikely, see Sect.\,\ref{hbe}), (2) it is present, but the orbit is very eccentric or the orbital period is larger than the time covered by our spectroscopic observations that prevented us from detecting any noticeable RV variations, or (3) the secondary companion is $\sim$10 or more times fainter in the optical region than the primary. In the latter case, HD\,85567 might be a system similar to MWC\,728 (but with a larger components'  mass ratio) or to CI\,Cam \citep[the only known FS\,CMa type binary with a degenerate secondary component, ][]{2006ARep...50..664B}. Note that the secondary companion in CI\,Cam revealed itself by the presence of a weak He {\sc ii} 4686 \AA\ line, which regularly moves in the spectrum with a period of $19\fd41$, and by an outburst in 1998, which was observed in from X-rays to radio waves. With different orbital parameters and undetected outbursts, a degenerate secondary component in HD\,85567 may remain unrevealed. To the best of our knowledge, the object has not been detected in X-ray surveys.

No signs of a direct binary interaction have been detected in the properties of HD\,85567. However, the presence of warm dust in its vicinity indicates that such an interaction might have taken place in the past. The 9.7--$\mu$m silicate feature in a Spitzer Space Observatory spectrum of HD\,85567 is seen in emission (see Fig.\,\ref{f9}) indicating that the dust is optically-thin. At the same time, the feature is broad that was interpreted by the presence of large grains of mostly olivine and pyroxene \citep[see][]{2012A&A...538A.101V}, implying that the dust has been exposed to the stellar radiation for some time.
 We also note that since the radiation from circumstellar dust dominates the object's SED at $\lambda = 2 \mu$m (see Fig.\,\ref{f9}), both \citet{2013A&A...558A.116W} and \citet{2014A&A...569A..25V} had a little chance to detect the possible secondary component in this spectral region.

\subsection{Circumstellar environments}\label{cs_env}

The emission-line profiles in the spectrum of HD\,85567 described in Sect.\,\ref{emission} indicate that the circumstellar matter distribution is quite
complicated. Most of the double-peaked Fe {\sc ii} and hydrogen Paschen series lines exhibit nearly equal emission peak strengths, the [O {\sc i}] lines show
the blue-shifted peak slightly stronger than the red-shifted one, and the Balmer lines show the red-shifted peak stronger than the blue-shifted one. The latter
also show blue-shifted absorption components and additional emission peaks, which appear near the line blue boundary (set by the absorption component at RVs
of $\sim -500$ km\,s$^{-1}$)  and travel toward the line center on time scale of days.  The Fe {\sc ii} lines also show profile variations, which are reminiscent of
those of the Balmer line profiles, while the Paschen line profiles are less variable. The emission peak separation is larger in the Fe {\sc ii} lines compared to that
of the Paschen lines and to the [O {\sc i}] lines (see Sect.\,\ref{emission} and Fig.\,\ref{f5}).

These features suggest that the [O {\sc i}] lines are formed further away from the star where the rotational velocity of the circumstellar material is much lower.
The Fe {\sc ii} lines and the Balmer lines share a formation region near the star, and their profiles reflect the matter density variations there. A smaller peak
separation in the Balmer lines is most likely due to a larger optical depth in them.
The Paschen lines are formed in an intermediate region, where the matter density is lower and its variations produce a very small effect on the line profiles.

The double-peaked profiles with no additional features are typically explained by a disk-like matter distribution dominated by rotation. The Balmer line profiles
can be understood in the framework of a qualitative model, such as that proposed by \citet{2015ApJ...809..129M} for MWC\,728. This object shows only a
double-peaked structure of the Balmer lines with a stronger red-shifted emission peak and no blue-shifted absorption. The model assumes a noticeable
contribution from a stellar wind responsible for depressing the blue-shifted emission peak. If the wind contribution is even stronger and its terminal velocity
is larger than the most negative velocity of the double-peaked structure, a blue-shifted absorption component may show up in the line profile.
However, this component was almost undetectable in a number of the spectra (e.g., 04/18/2015--04/27/2015, see left panel of Fig.\,\ref{f6} and Fig.\,\ref{f8}).

The profile short-term variations may be
due to ejection of a blob that propagates toward the observer with a gradually reducing velocity as it moves away from the star. Initially such a blob is
dense and moves through the blue-shifted absorption. Later, when the blob becomes less dense, it appears as a traveling additional emission. A transition from
absorption to emission was observed in our spectrum taken on 03/30/2012 (see  an additional blue-shifted emission peak at RV $\sim -150$ km\,s$^{-1}$ in Fig. \ref{f8}).
 Using the star's fundamental parameters (see Sect.\,\ref{parameters}) we determined its rotational period of $\sim 8\fd4$ and found that the Balmer line variability
is not connected with rotation of the star.

Our spectra also show long-term emission-line profile variations. The red emission peak intensity of the H$\alpha$ line gradually increases over time, while the blue peak intensity slightly decreases (see Table\,\ref{t5} and right panel of Fig.\,\ref{f7}). The relative increase of the red peak compared to the blue peak can be interpreted as a sign of an increasing amount of circumstellar gas in both the wind and disk regions. This suggestion can be further explored with line profile modeling, which is beyond the scope of this paper.

\section{Conclusions}\label{conclusions}

We presented the results of the high-resolution spectroscopic and multicolor photometric observations of the FS\,CMa type object HD\,85567. The main results of
our study of the object's spectrum include detection of a fast variability of emission-line profiles on a timescale of days, refinement of the fundamental parameters of
the underlying B--type star and the distance toward it (see  Table\,\ref{t1}), and finding that the absorption metallic lines (e.g., Mg {\sc ii}, Si {\sc ii}, see Table\,\ref{t3})
show very small position variations. The latter exceeds the measurement uncertainty ($\sim$0.1 km\,s$^{-1}$) but does not seem to be explained by the orbital motion
due to a small number of observations and possible contamination by the circumstellar material. We also found no support for the pre-main-sequence evolutionary
state of HD\,85567 as argued by \citet{2012A&A...538A.101V} and \citet{2013A&A...558A.116W}. There is no evidence of a star-forming region in its vicinity, and its position on the Hertzsprung-Russell diagram combined with the circumstellar dust distribution are inconsistent with a young age.

At the same time, we found no compelling evidence for the object's binarity. Its relatively strong emission-line spectrum and a large IR excess imply the presence of large amounts of circumstellar gas and dust that are not expected to exist around a nearly main-sequence single star of a relatively low mass ($\sim$6 M$_{\odot}$). A secondary companion may be too faint for its signature to be detected in the optical spectrum or have an eccentric orbit, so that our observations did not cover orbital phases near a periastron. Even if it is a binary, no signs of an active mass transfer between the components have been found. Explanation of the shape of the silicate emission feature at 9.7--$\mu$m requires some processing time by stellar radiation, thus supporting not a recent dust formation.

HD\,85567 is unlikely a result of a merger in a binary system. Its low rotation speed ($\sim$40 km\,s$^{-1}$, Sect.\,\ref{parameters}) would imply a significant angular momentum loss from the system. If this is the case, the event should have had occurred long ago, as no variations of the optical brightness has been detected during the last $\sim$40 years.

Further regular spectroscopic and photometric observations are needed to clarify the object's nature. High-resolution spectroscopy every few days during several observing seasons may solve the orbital motion problem and put more constraints on the circumstellar gas dynamics.

\acknowledgements S.K. acknowledges help from the University of North Carolina at Greensboro in organizing his trips to and stay in Greensboro, NC.
R.M. acknowledges support by VRID--Enlace 216.016.002--1.0 and the BASAL Centro de Astrof{\'{i}}sica y Tecnolog{\'{i}}as Afines (CATA) PFB--06/2007. We thank David Hollenbach for useful discussion of the photo-evaporation theory and its predictions.
 We thank Y.~Fremat for making a Fourier transfer code to measure rotational velocities available to us.
This research has made use of the SIMBAD database, operated at CDS, Strasbourg, France.
This publication makes use of data products from the Two Micron All Sky Survey, which is a joint project of the University of Massachusetts
and the Infrared Processing and Analysis Center/California Institute of Technology, funded by the National Aeronautics and Space Administration
and the National Science Foundation.

\bibliography{ms_rev}

\end{document}